\def\breakon{\end{multicols}\widetext\vspace{.5cm}  
\noindent\rule{.48\linewidth}{.3mm}\rule{.3mm}{.5cm}\vspace{.5cm}}  
\def\breakoff{\vspace{.5cm}  
\noindent  
\rule{.52\linewidth}{.0mm}\rule[-.47cm]{.3mm}{.5cm}\rule{.48\linewidth}{.3mm}  
\vspace{.5cm}  
\begin{multicols}{2}  
\narrowtext}

\documentclass[aps,draft,epsf,twocolumn,superscriptaddress,eqsecnum]{revtex4}
\input epsf
\newcommand{\be}{\begin{equation}}  
\newcommand{\ee}{\end{equation}}  
\newcommand{\bea}{\begin{eqnarray}}  
\newcommand{\eea}{\end{eqnarray}}

\begin{document}  
  
%\draft  
%%%%%%%%%%%%%%%%%%%%%%%%%%%%%%%%%%%%%%%%%%%%%%%%%%%%%%%  
\title{Theory of the Quantum Hall Smectic Phase, II: Microscopic Theory}  
%%%%%%%%%%%%%%%%%%%%%%%%%%%%%%%%%%%%%%%%%%%%%%%%%%%%%%%  
\author{Daniel G.\ Barci}  
\affiliation{Department of Physics, University of Illinois at  
Urbana-Champaign, 1110  
W.\ Green St.\ , Urbana, IL  61801-3080, USA}
\affiliation{ Departamento de F\'\i sica Te\'orica,  
Universidade do Estado do Rio de Janeiro, Rua S\~ao Francisco Xavier 524, 20550-  
013, Rio de Janeiro, RJ, Brazil}  
\altaffiliation[Permanent Address: ]{Departamento de F\'\i sica Te\'orica,
Universidade do Estado do Rio de Janeiro, Rua S\~ao Francisco Xavier 524, 20550-
013, Rio de Janeiro, RJ, Brazil}
\author{Eduardo Fradkin}
\affiliation{Department of Physics, University of Illinois at
Urbana-Champaign, 1110 W.\ Green St.\ , Urbana, IL  61801-3080, USA}
\date{\today}

%%%%%%%%%%%%%%%%%%%%%%%%%%%%%%%%%%%%%%%%%%%%%%%%%%%%%%%%%%%%%%%%%%%%%%%  
  
\begin{abstract}  
We present a microscopic derivation of the hydrodynamic theory of   
the Quantum Hall smectic or stripe phase of a two-dimensional   
electron gas in a large magnetic field. The effective action of the low energy   
is derived here from a microscopic picture by integrating out high energy   
excitations with a scale of the order the cyclotron energy.    
The remaining low-energy theory can be expressed in terms of   
two canonically conjugate sets of degrees of freedom:   
the displacement field, that describes the  
fluctuations of the shapes of the stripes, and the local charge fluctuations   
on each stripe.   
\end{abstract}  
\maketitle    
%\begin{multicols}{2}  
  
%\narrowtext  
  
%%%%%%%%%%%%%%%%%%%%%%%%%%%%%%%%%%%%%%%%%%%%%%%%%%%%%%%%%%%%%%%%%%%%%%%  
\section{Introduction}  
%%%%%%%%%%%%%%%%%%%%%%%%%%%%%%%%%%%%%%%%%%%%%%%%%%%%%%%%%%%%%%%%%%%%%%%  
  
It is by now generally accepted that electron correlations in a 2DEG at   
sufficiently high Landau levels, are responsible for the large anisotropies   
in the transport properties observed in recent experiments on extremely high   
mobility samples in large magnetic   
fields\cite{Lilly,du,subsequent}.  
Analyzing fluctuations around a  Hartree-Fock stripe state  
\cite{platzman,fogler,chalker,tudor},   
and exploiting an  
analogy with the stripe related phases of other strongly correlated electron   
systems\cite{nature},  
Fradkin and Kivelson proposed \cite{FK} that the ground states of Quantum Hall   
Systems with   
partially filled Landau levels with $N \ge 2$ are predominantly  
electronic liquid crystalline.   
  
In a separate paper, coauthored with S.\ A.\ Kivelson and V.\ Oganesyan 
\cite{BFKO}, hereafter referred to as paper I, we proposed
an effective low-energy theory for the Quantum Hall Smectic described in terms of   
the Goldstone modes of the broken translation and rotation symmetry.  
The effective low energy Lagrangian for this state
is given by   
\be  
{\cal L}= \frac {eB} {\lambda} u \partial_t \phi - \frac {\kappa_{\parallel}} 2   
(\partial_x\phi)^2 - \frac {\kappa_{\perp}} 2 ( \partial_y u)^2 -\frac Q 2  
(\partial_x^2 u)^2.  
\label{eq:Lsm}  
\ee 
where we have assumed that the stripes run along the $x$ direction. 
Here $u(x,y)$ is the displacement field, the Goldstone mode of the broken  
translation symmetry, representing the transverse displacements of the
stripes 
({\textit i.\ e.\/ } along the $y$ direction); $\phi$ is the Luttinger field, representing  
the charge fluctuations on each stripe; $\lambda$ is the wavelength of the stripe  
state; $\kappa_\parallel$, $\kappa_\perp$ and $Q$ are elastic constants. 

In this paper we give a physically intuitive microscopic derivation of
the effective low-energy theory of Eq.\ \ref{eq:Lsm}. The approach that we will use here
is based on a microscopic theory  
which focuses on the role of dynamical
shape fluctuations in the quantum Hall stripe state and to 
their coupling with the charge fluctuations in this state. Thus, we will pay
special attention to the role played by the displacement field $u$, as well as to its coupling to the charge fluctuations on each stripe
represented by the Luttinger field $\phi$.  

\begin{figure}
\begin{center}
\leavevmode
%\vspace{.2cm} 
\noindent
%\hspace{1.0 in}
\epsfxsize=7 cm
\epsfysize=7 cm
\epsfbox{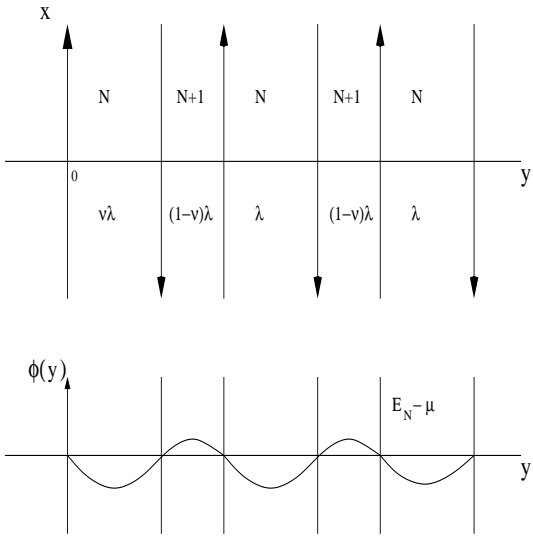}
%\vspace{.5cm}
\end{center}
\caption
{Schematic representation of the quantum Hall smectic state as a set of stripes, running
along the vertical $x$ axis. Here $N+\nu^*$ is the total filling factor, 
$\lambda$ is the period of the stripe, $\phi(y)$
is the effective local potential and $\mu$ is the effective chemical potential. The
arrows represent the internal chiral edge states. }
\label{figstripe}
\end{figure}

The construction that we use here is partially inspired by the picture of the quantum Hall smectic suggested by Hartree-Fock
calculations\cite{platzman,fogler,chalker,tudor,fertig,fertig-stability,fertig-cote}. 
Koulakov and coworkers\cite{fogler} and Moessner and Chalker\cite{chalker} found  
that the stripe state can be viewed as a set of $N$ filled Landau levels, with a charge  
modulation due to the electrons in the partially filled Landau level (see figure \ref{figstripe}).  
For fully polarized (``spinless") fermions at a static level this state looks like an array of strips of charge,  
corresponding to regions with an effective filling factor $N+1$, surrounded by regions with  
filling factor $N$. Thus, the electrons arrange themselves in a state which  
locally mimics a gapped integer Hall state. If this charge-modulated  state was due to an imposed external potential,  
inside these regions the electron fluid would be incompressible and only the excitations  
at their ``edges" would remain gapless. Hence, at a static level,  the state looks like an  
array of fixed chiral Luttinger liquids, the edges of integer quantum Hall stripes, a picture advocated by Fradkin and Kivelson \cite{FK}, and by MacDonald and Fisher\cite{MF} (see also ref.\ \cite{EFKL,carpentier}). This, however, is not the full story since this charge modulated state is a self-consistent ground state, and not the result on any externally applied potential.

In an insightful paper, MacDonald and Fisher investigated the properties of the quantum Hall
smectic viewed as an array of coupled chiral Luttinger liquids subject to constraints
imposed by the requirement of global rotational invariance. One of the issues raised in
ref.\ \cite{MF} is how many independent degrees of freedom does the quantum Hall smectic
actually have. Fradkin and Kivelson\cite{FK} had advocated a picture in which both
charge and shape degrees of freedom although coupled were both 
part of the physical picture. Direct inspection of the effective action of Eq.\
\ref{eq:Lsm} shows that the displacement field $u$, which embodies the shape fluctuations,
and the Luttinger field $\phi$ of charge fluctuations are canonically conjugate dynamical
variable much as coordinates and momenta are in Classical Mechanics. This connection is a
direct manifestation of the Lorentz force, crucial for the dynamics of charged particles
in magnetic fields.  Thus, although shape and charge are both useful descriptions of the
physics, we find that they are not truly independent degrees of freedom, in agreement with the
point of view of MacDonald and Fisher.

A simple change of basis relates the quantum smectic picture, which uses as degrees of freedom
 the displacement field $u$ and the Luttinger field $\phi$, and  the picture of an array of 
coupled chiral Luttinger liquids:
\bea 
u&\to&\partial_x\phi^++\partial_x\phi^-\\ 
\phi&\to&\phi^+-\phi^- 
\eea 
In the chiral Luttinger liquid basis the effective Lagrangian of Eq.\ (\ref{eq:Lsm}) theory takes the same form  
as in the effective theory of MacDonald and Fisher\cite{MF}.

%%%%%%%%%%%%%%%%%%%%%%%%%%%%%%%%%%%%%%%%%%%%%%%%%%%%%%%%%%%%%%%%%%%%%%%%%%% 
%%%%%%%%%%%%%%%%%%%%%%%%%%%%%%%%%%%%%%%%%%%%%%%%%%%%%%%%%%%%%%%%%%%%%%%%%%%    

In Hartree-Fock theories of the stripe state\cite{fogler,chalker,fertig,fertig-cote,fertig-stability,tudor}, the stripe state is 
derived by balancing the energy density of the Hartree contribution, which favors phase 
separation, against the energy density of the Fock contribution which, as an exchange driven 
effect favors the spreading out of the charge. In these calculations, Landau level mixing is taken into account in the form of an effective interaction for electrons in a partially filled Landau level but it is otherwise ignored.The resulting stripe state is structure with a fairly short wavelength $\lambda$. For Landau level index $N=2$, Hartree-Fock calculations yield a wavelength  $\lambda \sim 3 \ell$, where $\ell$ is the magnetic length, whereas for $N \gg 1$, Moessner and Chalker find~\cite{chalker} $\lambda \sim \textrm{const.}\; \sqrt{N} \ell$, where the constant factor is a number of order $3$. At the level of these calculations~\cite{fogler,chalker} dynamics is ignored. Dynamics is incorporated later on, at the level of a time-dependent Hartree-Fock approximation, as in the extensive work of Fertig and coworkers~\cite{fertig-cote,fertig-stability}. Using these methods Fertig and coworkers calculated the spectrum of collective modes, and studied the problem of the stability of the stripe state with respect to a possible stripe crystal. However, while Hartree-Fock theories have been very successful in predicting a smectic state and in studying some of its important properties, they do not yield a transparent picture of the role of the quantum fluctuations of the edges of the charge stripes. Conceptually, this is an undesirable feature since the displacement fields of these edges are the Goldstone bosons of the spontaneously broken translation symmetry.

The main purpose of this paper is to construct an effective theory in which the effects and the dynamics of the displacement field $u$, the Goldstone boson of the broken translation symmetry of the quantum Hall smectic, and of the Luttinger field $\phi$, is made explicit. Although it is possible to derive
the effective theory by a detailed analysis of a Hartree-Fock calculation, as was done
by Lopatnikova and coworkers\cite{bert} recently and independently from this work, here we will introduce instead an alternative
approach based on the intuitive picture of the stripe state as an array of regions of integer quantum Hall states separated by dynamical edges. 
For simplicity we will consider the case of short range interactions with a pair potential given in Eq.\ \ref{eq:potential}, instead of a long range Coulomb interaction. Although long range interactions change the behavior of the charged collective modes, their precise form are not central neither to the existence of the stripe state itself nor to much (but not all) of the qualitative physics.

We begin by constructing an effective local potential by means of a decoupling of the microscopic density-density 
interactions in terms of a Hubbard-Stratonovich field $\varphi$, and then proceeding to (formally) 
integrating out the fermionic degrees of freedom.  At the level of a static approximation, 
the field $\varphi$ plays essentially the role of a scalar (Hartree) potential  
self-consistently generated by the electron-electron interactions. For  a system of electrons in a fixed area, the Hartree term leads to a phase-separation  instability. As we pointed out above, the exchange effects of the Fock term stabilize a stripe structure. We will show here that it is possible to stabilize the stripe state instead by the quantum fluctuations of the edge states and by a suitable choice of boundary conditions.
Below we construct an effective action which includes the effects of the Hartree contributions and of 
the quantum fluctuations of the edge states and show that this action leads to a stable stripe solution provided we choose boundary conditions with a fixed number of stripe wavelengths.  The resulting stripe state that we find has a wavelength comparable in scale to the wavelength found by Hartree-Fock calculations~\cite{fogler,chalker}, but the ground state energy is not as good. (Naturally, it is possible to compute the effects of the Fock terms.)
Nevertheless, the approach use here leads to  an effective action parametrized  
instead by the geometry of the stripe states, {\it i.\ e.\/} the positions of the stripes (or ``internal edges"),  
and by the local charge fluctuations on each stripe. Furthermore, we also find that the elastic constants of the effective theory are physically sensible.

The next step is to construct the stripe ground state as a solution of the  
saddle-point (self-consistent) equations derived from an effective action as described above. 
We will assume that the stripe state locally looks like regions of a integer quantum Hall  states separated by edges.  
We then construct the stripe solution with a fixed integer number $n$ of periods  
of wavelength $\lambda$ for the stripe for a geometry of width $L_y=n\lambda$.  
We will then compute the total energy of this state with a fixed number of periods. This 
total energy has two contributions: (a) a piece coming from the ``bulk regions" and (b) a 
piece coming from the ``edges". We will then find the optimal 
solution by minimizing the total energy  
as the period $\lambda$ is varied while keeping the number of periods $n$ fixed.  

The saddle-point solution thus constructed, $\varphi_{sp}$, is a smooth differentiable 
function of the coordinate $y$ normal to the stripe orientation. It varies 
periodically in space about an average value which plays the role of the Fermi  
energy. The plane  is thus split
into two types of regions: (a) regions in which the field $\varphi_{sp}$ is nearly  
constant and far from the Fermi 
energy, (b) and  regions in which $\varphi_{sp}$ crosses the Fermi energy.  
In the former, the system behaves as a perturbed Landau level problem with a  
full gap in the single particle 
spectrum. Inside these regions we can approximate the effective action  
for slow time and space dependent fluctuations of $\varphi$ by means a gradient  
expansion of the fermionic determinant. In other terms, we will keep field configurations which vary slowly on the scale of the cyclotron gap and are smooth on scales long compared with the cyclotron length. This procedure is safe away from edge states. However, wherever $\varphi_{sp}$ crosses 
the Fermi level, the system has gapless fermionic excitations. In these regions 
it behaves like an edge state with a definite Fermi velocity determined by the slope of 
the stripe solution. These edge states regions must be taken into account explicitly  
in order stabilize the stripe state. (These approximations are very accurate if the wavelength of the stripe state is long compared to the cyclotron length. However, for reasonable interactions it is not the case. Nonetheless we will work within this approximation since it yields qualitatively correct answers.) 
 
Thus, instead of regarding the edges as quasi-static structures,  
the quantum Hall smectic is a theory of fermions moving on {\em fluctuating  
stripes}. This motion of the stripes is physically due to the fact that the position  
of the stripes is defined arbitrarily (up to a displacement by an integer number of wavelengths). 
This arbitrariness is due to the fact that the translation symmetry normal to the  
stripe is spontaneously broken in this state. Since this is a continuous symmetry, there should be a Goldstone  
boson associated with it, which we will parametrize by  
the displacement field $u$ of the stripe position. In other terms, a 
correct quantum theory of this state requires that these collective modes be quantized 
correctly. We will do this by parametrizing the physically important  
configurations as deformed stripe solutions $\varphi_{sp}$ of the form 
\begin{equation} 
\varphi_{sp}=\varphi_{sp}\left(y \sqrt{1-\frac{1}{2}(\partial_x u)^2}+u\right) 
\label{eq:parametrization} 
\end{equation} 
where $u=u(x,y,t)$ is the displacement field. Hence, we will describe the  
fluctuations of the quantum hall smectic in terms of two sets of degrees of freedom:   
(a) the ``internal'' chiral fermions and (b) the shape fluctuations  
represented by a dynamical displacement field. Nevertheless, we will see below that these degrees  
of freedom are not independent from each other and that they are related  in a manner dictated entirely by 
the quantum mechanics of a charged fluid in a magnetic field. 
 
This paper is organized as follows. In Section \ref{sec:Model} we present the derivation of the effective  
action for the stripe state. Here we construct an effective action for 
$\varphi$ in the ``bulk regions" and at long wavelengths, and show how it couples  
to the edge modes. These in turn will be written in terms of a set of chiral  edge bosons. 
We will use this effective action to find the optimal 
stripe solution of the saddle-point equations, which is presented in Section \ref{saddlepoint}.  In this section, and in Appendix \ref{sec:mean-field}, we discuss how the (Hartree) stripe solution is stabilized by quantum edge fluctuations.
It turns out that the results that follow from the solution that we will derive here has the same physical 
properties and it is for all practical purposes equivalent to the results of the Hartree-Fock theories. 
In Section \ref{fluctuations} we analyze the effect of quantum fluctuations around the saddle-point.  
Here we derive the coupling between the displacement field $u$ and the non-chiral Luttinger  
field $\phi$. In this section we give an estimate of the elastic constants entering in  Eq.\ (\ref{eq:Lsm}). 
Finally, in Section \ref{conclusions} we discuss our results and present our conclusions.  
In the Appendices we give technical details of the derivations discussed in the text. 
 
%%%%%%%%%%%%%%%%%%%%%%%%%%%%%%%%%%%%%%%%%%%%%%%%%%%%%%%%%%%%%%%%%%%%%%% 
\section{Effective action for the quantum Hall stripe state} 
 
%%%%%%%%%%%%%%%%%%%%%%%%%%%%%%%%%%%%%%%%%%%%%%%%%%%%%%%%%%%%%%%%%%%%%%% 
\label{sec:Model} 
 
In this section we will derive an effective action well suited for description of an 
inhomogeneous state such as a stripe state. Thus we will begin with a microscopic 
theory of interacting electrons in a large magnetic field and identify the  
degrees of 
freedom needed to construct a stripe state. In section \ref{saddlepoint} 
we will find the optimal stripe solution by means of a variational approach.  
 
The generating functional of two-dimensional 
(non-relativistic) interacting fully spin-polarized electrons in a 
perpendicular magnetic field is 
\be 
Z[{\tilde A}_\mu]=\int {\cal D}\psi^*{\cal D}\psi~e^{i S(\psi,{\tilde A}_\mu)} 
\ee 
where 
\bea 
S&=&\int d^3z~ \left[\psi^*(z)[iD_0-\mu]\psi(z) 
+\frac{1}{2m} 
\left|{\bf D}\psi(z)\right|^2\right] \nonumber \\ 
&-&\frac{e^2}{2}\int d^3z d^3z'~|\psi(z)|^2V(z-z')|\psi(z')|^2 
\label{model} 
\eea 
and 
\be 
D_\mu=\partial_\mu+ie A_\mu +ie {\tilde A}_\mu 
\ee 
is the covariant derivative.  
Here $\vec\nabla\times\vec A={\vec B}$ is the external uniform magnetic field, 
$V(x-y)$ is a two-body interaction potential and ${\tilde A}_\mu$ are small 
electromagnetic perturbations introduced to probe the system. 
 
We can decouple the quartic interaction term by means of a Hubbard-Stratonovich 
(HS) transformation, and introduce a new field $\varphi$. The generating 
functional now takes the form, 
\be 
Z[{\tilde A}_\mu]=\int {\cal D}\psi^*{\cal D}\psi{\cal D}\varphi~ 
e^{ i S[\psi,{\tilde A}_\mu,\varphi]} 
\ee 
with 
\bea 
S&=&\int d^3z~ \left[\psi^*(z)[iD_0-\mu- \varphi]\psi(z) 
+\frac{1}{2m} 
\left|{\bf D}\psi(z)\right|^2\right] \nonumber \\ 
&+& \frac{1}{2}\int d^3z \int d^3z'~\varphi(z)V^{-1}(z-z')\varphi(z') 
\eea 
where $V^{-1}$ is the inverse of the instantaneous pair potential operator  
$V(z-z')$. 
Much of what we will discuss here can be done for any pair potential.  
However, to be able to find an  explicit analytic solution 
we will work with a short range interaction with 
coupling constant $g$ and range $a$. In any event at this level the physics will not 
depend too much on the details of the pair interaction. 
In particular, there exists a choice of short range pair interactions for which the 
contribution of the interaction term to the action reduces to the following local 
expression: 
\bea 
\lefteqn{\frac{1}{2}\int \; d^3x \int \; d^3x' \; 
\varphi(x)V^{-1}(x-x')\varphi(x') 
}  &&\nonumber \\ 
&&\equiv \int d^3x \;  \left( \; {\frac{a^2}{2g}} \left(\nabla \varphi(x) 
\right)^2 
 +{\frac{1}{2g}}  \varphi(x)^2\right) 
\eea 
provided that $V(\vec x)$ is the short ranged interaction potential 
\be 
V({\vec x})={\frac{g}{2\pi a^2}} 
K_0\left({\frac{ |{\vec x}|}{a}}\right) 
\label{eq:potential}
\ee 
where $K_0(z)$ is the modified Bessel function. 
 
The fermionic action is now quadratic and, formally, the fermionic path 
integral can be carried out obtaining 
\be 
Z({\tilde A}_\mu)=\int{\cal D}\varphi  
e^{i S_{\rm eff}[{\tilde A}_\mu,\varphi]} 
\label{Z} 
\ee 
where $S_{\rm eff}({\tilde A}_\mu\varphi)$ is given by 
\bea 
S_{\rm eff}&=&-i {\rm Tr} \ln\left[iD_0-\mu- \varphi 
+\frac{1}{2m}{\bf D}^2\right] \nonumber \\ 
&+&\frac{1}{2}\int d^3z d^3z'~\varphi(z)V^{-1}(z-z')\varphi(z') 
\label{Seffexacta} 
\eea 
This effective action is well defined provided the fermion determinant does not have 
any zero eigenvalues.  
However, even for fairly general smooth configurations of the field $\varphi$ 
there can be zero modes in the fermion determinant. To see this, let us consider 
configurations in which $\varphi$ varies very slowly.  
In this case, the main effect of the field $\varphi$ is to shift the single particle 
energies of the electrons, {\it i.\ e.\/} the energies of the Landau levels will vary 
from point to point but sufficiently slowly  so that Landau level mixing can be 
ignored to a first approximation. Thus, at least locally in space, 
the electrons fill up an integer number $N$ of Landau levels. Clearly, the 
Hubbard-Stratonovich field $\varphi$ plays the role of an 
effective local chemical potential. 
Thus, almost everywhere in space, for these field configurations  
there is a gap in the fermionic spectrum of the   
order of the cyclotron energy, $\hbar\omega_{c}$. In this case, the 
fermion determinant is well behaved. It is well known that, 
for such regular configurations, the effective action, $S_{\rm bulk}$,  
is a local functional of 
$\varphi$ and its derivatives\cite{sakita,hosotani,salam,LF,BLA}. 
However, where 
$\varphi$ crosses the Fermi energy, {\it i.e.} where the number of  
filled Landau levels changes from $N$ to $N-1$, the gap vanishes, 
there are fermion zero modes, and hence the fermion determinant is  
badly behaved.   
The points of the plane where, at fixed time $t$, $\varphi$ crosses 
the chemical potential define a set of instantaneous curves (or ``strings'').  
These curves are internal ``edges'' that enclose regions with a given  
integer filling factor.  
 
Therefore, instead of blindly integrating out all the fermionic modes, we 
will integrate out all modes with energy greater than or of order $\hbar  
\omega_{c}$. For a system with filling factor $\nu=N+\nu^*$, where $\nu^*$ is the 
effective filling factor of the partially filled Landau level, we can indeed integrate 
out all fermionic states without difficulty except for those states on the $N$-th 
Landau level with support on the ``strings". Thus we will treat these states 
separately. We will see below that these states will play a crucial role in the 
dynamics of the quantum Hall stripe state. Thus, the full system can be described 
in terms of an effective action of the form 
\be 
S_{\rm eff}[\varphi,\tilde\psi]=S_{\rm bulk}[\varphi]+ 
S_{\rm string}[\varphi,\tilde\psi] 
\label{completeaction} 
\ee 
where $S_{\rm bulk}[\varphi]$ is the effective action of the of the field $\varphi$ due to 
both the bulk regions and to the filled Landau levels.  
In eq.\ \ref{completeaction}, $S_{\rm string}$ is the contribution to the action  
due to the low energy (chiral) fermion modes, $\tilde \psi$,  
localized in the neighborhood of 
the strings, which have not been integrated out.  Note that both the  
position of the strings and the effective edge-potential seen by the 
chiral fermions are implicit functions of the field configuration,  
$\varphi$. 
In general the strings are {\sl dynamical}, 
with a non trivial time dependence which has to be included explicitly in the 
path-integral. In addition, although at the level of the bare Hamiltonian the 
field $\varphi$ is a space and time-dependent field with 
no independent dynamics of its own, the fluctuations of the bulk 
regions, {\it i.\ e.\/} the regions where the filling factor is constant, 
induce non-trivial dynamics for the field $\varphi$.  
We will show below and  
in Appendix \ref{sec:CH} that the necessity of retaining the chiral  
fermion zero modes along the strings (and much about the form of  
$S_{\rm string}$) could be deduced, even were we  
have blindly integrated out all the fermionic modes, from the  
requirements of gauge  invariance. 
 
By definition, the effective action $S_{\rm bulk}[\varphi]$ can be constructed  
perturbatively 
as the sum of all the one-particle irreducible correlation functions of the field 
$\varphi$ (see for instance \cite{book}). The procedure outlined in Eq.\ 
\ref{Seffexacta} yields the one-loop approximation to $S_{\rm bulk}[\varphi]$, {\it 
i.\ e.\/} this is the Hartree approximation with RPA corrections. To lowest order, the one-loop approximation yields the contribution to the effective action from particle-hole fluctuations between the topmost occupied Landau level and the first unoccupied Landau level. This is the contribution with the leading residue at long wavelengths. There are other one-loop contributions but have smaller residue and larger energy denominators. Thus, in the leading order in $1/B$, in which Landau level mixing is not taken into account, only the term with the leading residue are important. In this paper we will only keep the contributions from the leading residue since they are the largest and play and saturate the sum rules (at low momenta). The typical form of these terms can be found, for instance, in ref.\ \cite{LF}. There it is shown that, in momentum space,  these terms have residue proportional to $\vec q^2$, dictated by current conservation, multiplied by a Laguerre polynomial of the variable $\vec q^2/B$, and an energy pole at the cyclotron frequency. In this paper we will make the (crude) approximation of setting both the energy denominator and the Laguerre polynomial at their zero frequency and zero momentum values. The approximate form of the effective action that results is accurate for long wavelengths and for slowly varying excitations. We will find below that the wavelength of the stripe state is a actually not long (in fact about 3 magnetic lengths) and hence this approximation is not accurate.  Nevertheless it does yield a number of qualitatively correct results. We have checked, for instance, that including the full frequency dependence does not appreciably change our  results. Thus, for the sake of simplicity we will use the long wavelength, low frequency approximation. However, this approximation does include the effects of the quantum fluctuations of the ``internal edges" which will play an important role here.

In principle it is 
straightforward, but tedious, to add further corrections to $S_{\rm bulk}[\varphi]$, such as  the Fock or exchange
correction, which plays a crucial role in 
stabilizing the stripe solution\cite{fogler,chalker,bert}. However we will find in 
section \ref{saddlepoint} that by a suitable choice of boundary conditions it is 
possible to stabilize the stripe state with the contributions from the quantum fluctuations of the ``internal edges", without 
including the Fock terms. Although energetically the results found at this level of approximation are not as good energetically as in 
Hartree-Fock the solution, this procedure turns out to yield a state which, at least qualitatively, has very similar properties to the one found in Hartree-Fock. Thus, for instance, we will find that the wavelength of the stripe state is very close to the Hartree-Fock results~\cite{fogler}. In the remainder of this paper we will use the one-loop approximation to $S_{\rm bulk}[\varphi]$. 
 
\subsection{Contributions from incompressible regions} 
\label{sec:bulk} 
 
The form of $S_{\rm bulk}$  
can be computed quite easily. It has essentially the same  
form\cite{sakita,LF,BLA} as the effective 
action for weak and slowly varying electromagnetic perturbations in the 
integer quantum Hall effect:  
\be 
S_{\rm bulk}=S_{\varphi} + S_{A} 
\label{bulk} 
\ee 
where $S_{\varphi}$ is the one-loop effective action for the 
Hubbard-Stratonovich field $\varphi$  
\bea 
S_\varphi&=&\int d^3x 
\left\{\frac{\gamma^2(\varphi)}{4\pi\omega_c}\frac{3}{8 eB}  
\left(\nabla^2 \varphi \right)^2+ 
\left(\frac{\gamma(\varphi)}{4\pi\omega_c}+  
\frac{a^2}{2g}\right)\left(\nabla \varphi \right)^2\right. \nonumber \\ 
 &+&\frac{\varphi^2(x)}{2g} 
+ \left.  \frac{e}{2\pi} B \gamma(\varphi) 
\varphi(x)-\frac{e^2}{4\pi m}\gamma^2(\varphi) B^2 \right\} . 
\label{Svarphi} 
\eea 
The coupling to a weak electromagnetic perturbation ${\tilde A}_\mu$ yields 
the additional term in the effective action: 
\bea 
&&\lefteqn{S_A= 
\int d^3x~ \left\{ \frac{e^2}{2\pi} 
\gamma(\varphi) {\tilde A}_0 B+\frac{e^2}{4\pi} \gamma(\varphi) 
\epsilon^{\mu\nu\rho}{\tilde A}_\mu \partial_\nu {\tilde A}_\rho \right.} 
\nonumber \\ 
&&+\left. \frac{e^2 \gamma(\varphi)}{4\pi \omega_c}  {\bf \cal E}^2- 
\frac{e^2\gamma^2(\varphi)}{4\pi m}{\bf \cal B}^2 
+\frac{e \gamma(\varphi)}{2\pi}  {\bf \cal B} \varphi + 
\frac{e \gamma(\varphi)}{2\pi\omega_c} 
 \vec{\cal E}\cdot \vec\nabla\varphi \right\}\nonumber \\ 
&& 
\label{Sa} 
\eea 
In Eq.\ \ref{Svarphi} and Eq.\ \ref{Sa} we have neglected terms higher in 
derivatives and higher powers of the gauge field. These terms  
are functions of higher powers (and higher derivatives) of ${\cal E}$, ${\cal B}$ and $\varphi$. 
The coefficients of these terms are suppressed by higher powers of $1/B$ 
(see Ref.\ \cite{LF}). 
%These non-linear terms give rise to (uninteresting) subleading corrections.  
In Eq.\ \ref{Svarphi} and Eq.\ 
\ref{Sa} we have denoted by 
$\gamma(\varphi)$ the integer-valued function given by 
\begin{equation} 
\gamma(\varphi)=\sum_{n=0}^\infty \Theta\left(\mu+\varphi-(n+\frac{1}{2}) 
\omega_c\right) 
\label{gamma} 
\end{equation} 
where $\Theta(x)$ is the step function. Here 
$\gamma(\varphi)$ is an integer-valued function of the field $\varphi$, 
and counts the number of filled Landau levels. It jumps by one unit 
wherever $\varphi$ crosses the Fermi energy. 
Eq.\ (\ref{Sa}) represents the action of the small electromagnetic 
perturbations. The first term (linear in the perturbation) yields 
the constraint between the total charge 
density and the magnetic field ($\rho \propto B$). The other terms, quadratic 
in the 
electromagnetic perturbation, are 
Maxwell-Chern-Simons terms with a local Hall conductance 
for a system with an integer number of completely filled landau levels, 
given here by $\gamma(\varphi)$, an effective local dielectric 
constant $\varepsilon=e^2\gamma(\varphi)/(2\pi\omega_c)$, 
and an effective local magnetic permeability 
$\chi=e^2\gamma^2(\varphi)/(2\pi m)$. 
 
The effective action of Eq.\ \ref{Svarphi} gives an accurate description of the 
physics at distances long compared with the magnetic length and for frequencies low compared with the cyclotron  
frequency 
$\omega_c$. Thus, as it stands, this effective action does not describe the Kohn  
mode. To restore the 
effects of this collective mode it is necessary to consider the full density- 
density 
correlation function, which contains all even powers in the frequency\cite{LF}.  
We will ignore these effects since the degrees of freedom involved in the stripe  
state are concentrated at energies much less that $\hbar\omega_c$ 
and are decoupled 
from the Kohn mode. 
 
\subsection{Contributions from internal edges} 
\label{sec:edges} 
 
As a consequence of gauge invariance, the action Eq.\ \ref{Svarphi} 
does not contain terms with an explicit dependence on the time 
derivative of $\varphi$. Thus  it may seem that the 
field $\varphi$ has no independent dynamics in this 
approximation. However, the dynamics of $\varphi$ arises from the 
non-trivial physics associated with the strings defined by the 
discontinuities of $\gamma(\varphi)$.  
 
More specifically, we are interested in the case in which the system has $N$ 
completely filled Landau levels and the $N$-th Landau level is only partially 
filled. Then, on the points of the plane where 
\be 
\varphi(x,y,t)=\left(N+\frac{1}{2}\right)\omega_c-\mu 
\label{equipotential} 
\ee 
the gap collapses. Notice that Eq.\ \ref{equipotential} is just the 
argument of the function $\gamma(\varphi)$. 
Eq.\ \ref{equipotential} defines a generic time-dependent curve,  
$\vec R_n(s,t)$, where $n$ labels the string and $s$ is the arclength 
along the string. 
 
As discussed in Appendix \ref{sec:geometry}, 
for a general field configuration, $\varphi$, $S_{\rm 
string}$ is very complicated.   However, to the extent 
that the relevant configurations of 
$\varphi$ are quasi-static and smooth, this problem looks very similar to 
the standard edge  state  
problem. The main difference is that in the problem we are interested in here 
the electric  field 
which generates the ``edge" is itself a dynamical field. 
It is well known from the theory of edge states\cite{edgestates} 
in the IQHE, that in order to define a consistent, gauge-invariant, 
effective action for this system, it is necessary to add to the bulk action 
Eq.\ (\ref{bulk}), the action of one-dimensional chiral fermions (or its 
bosonized version) with support at the edge\cite{wen,frohlich,Kao,dror}. 
For an array of parallel static straight edges ({\it i.e.} for at a smectic 
saddle-point configuration, $\varphi_{sp}$,  
the bosonized effective action is 
\cite{wen} 
\be 
S_{\rm string}[\varphi_{sp},\phi]=\sum_n\int  \frac{ds dt} {4\pi}~  
\left\{\partial_t\phi_{n,\pm}\partial_s\phi_n\mp v 
\partial_s\phi_n\partial_s\phi_n \right\} 
\label{Schiral0} 
\ee 
where $v$ is the velocity of the chiral Bose fields $\phi_{\pm}$.  
The  
velocity 
$v$ is the sum of two contributions: (a) the drift velocity $c |{\vec \nabla} 
\varphi|/B$, where ${\vec \nabla} \varphi$ is the effective electric field  
normal to the 
edge, and (b) a finite renormalization due to the forward scattering  
interactions among the edge 
fermions. For a system with many edges there is also a host of possible forward  
scattering  
interactions 
that mix the edge states\cite{MF}. We will discuss these interaction below.  
The sign $\pm$ in Eq.\ \ref{Schiral0} is the chirality of each edge. 
 
It is also relatively straightforward, as shown in Appendix \ref{sec:geometry},  
to  
treat field configurations which represent small fluctuations about  
$\varphi_{sp}$. 
Between two nearby 
edges the quantum Hall fluid is incompressible. 
Thus, as the field 
$\varphi$ fluctuates it induces a charge redistribution at the edges. 
Physically this means that the edge fermions (and the equivalent chiral 
bosons) feel an effective dynamical longitudinal electric field due to 
the fluctuating geometry of the edges induced by the fluctuations of $\varphi$.  
The result is, to leading order in $|\varphi-\varphi_{sp}|$, 
\bea 
S_I&[\varphi,\phi]&\equiv S_{\rm string}[\varphi,\phi] -  S_{\rm 
string}[\varphi_{sp},\phi] 
\eea 
where 
\bea 
S_I&=&\sum_n\int dsdt~ \left\{ \varphi_{n,\pm} \; \partial_s\phi_{n,\pm} \right. 
\nonumber  \\ 
 &+& \frac{1}{v} 
\left( 
\partial_s R_{ni} 
\partial_t^2 R_{ni}\pm v 
\partial_s^2 R_{ni}  
\partial_t   R_{ni} 
\right)_\pm \;  
\partial_s\phi_{n,\pm} \nonumber \\ 
&+& 
\left( 
{\tilde A}_0+{\tilde A}_i  \left[ 
\partial_s R_{ni} 
\pm \partial_t R_{ni} \right] \right)_\pm \partial_s\phi_{n,\pm} 
\left. \right\} 
\label{interaction} 
\eea 
where $n$ labels the stripe, $\pm$ are the right and left moving edges of each  
stripe, 
 and $i=1,2$ are the components of the displacement  
vector $\vec R$.  
The first line of Eq.\ \ref{interaction} is the coupling 
of the instantaneous charge fluctuations with the local potential. Here  
$\varphi_n$ 
represents the fluctuating component of the Hubbard-Stratonovich field at the  
$n$-th 
stripe. The main effect of this term is to generate the conventional 
density-density interactions. The second line in Eq.\ \ref{interaction} 
is the geometrical coupling due to the {\sl time dependence} of the position on 
the $n$-th stripe $\vec R_n(s,t)$, or rather its {\sl displacement} away from  
the static mean 
field configuration. The last line represents the coupling of the 
$n$-th dynamical edge to an external electromagnetic perturbation. 
Notice that the coupling to the fluctuating geometry of the stripe has the same  
form 
as the coupling to a gauge field. 
 
The relation between the stripe displacement field $\vec R_n(s,t)$ and the 
Hubbard-Stratonovich field $\varphi$ can be found by differentiating Eq.\ 
\ref{equipotential} with respect to $s$ and $t$: 
\bea 
\frac{\partial \vec R}{\partial s}\cdot \vec\nabla \varphi &=& 0 
\label{relation1}\\ 
\frac{\partial \vec R}{\partial t}\cdot \vec\nabla \varphi &=& 
 -\frac{\partial \varphi}{\partial t } 
\label{relation2} 
\eea 
The interpretation of these equations is simple: Eq.\ (\ref{relation1}) 
tells us that, since the curve $\vec R(s,t)$ is an equipotential, 
then $\vec \nabla \varphi$ is normal to the edge direction $\partial \vec  
R/\partial s$. 
Eq.\ \ref{relation2} implies that 
the time variation of $\varphi$ produces a  
variation of $\vec R$ perpendicular  
to the curve.   
 
In what follows we will be interested primarily in the long wavelength  
fluctuations of the 
shapes of the stripes. In this regime, we need to keep track only of the  
fluctuations of 
${\vec R}_n$ normal to the stripe  (which will be considered to be straight on  
average). 
We will denote the normal component of ${\vec R}_n$ by the displacement field  
$u_n$. 
We will show in Section \ref{fluctuations} that the natural parametrization 
of the long wavelength fluctuations of the smectic phase by has the form 
\be 
\varphi=\varphi_{sp}[u]+\delta \varphi 
\ee 
where $\varphi_{sp}[u]$ is a solution of the saddle point equations {\sl locally  
deformed} (for the 
$n$-th stripe ) by the {\sl displacement field} $u_n(x)$, and $\delta \varphi$  
are the  
fluctuations of the gapped degrees of freedom. We will also find that the only  
role of the 
 geometric couplings of Eq.\ \ref{interaction}  is to renormalize the  
effective couplings. 
 In contrast, the first term of Eq.\ \ref{interaction} is ultimately responsible  
for the 
 dynamics of the smectic phase. From now on we will refer to $u_n$ as the 
 displacement field of the $n$-th stripe. 
A key property of the action $S_{\rm eff}[\varphi,\{\phi\}]$ 
is the way gauge invariance is realized in this phase: 
neither $S_{\rm bulk}$ nor $S_{\rm stripe}$ are separately gauge invariant, but 
their sum is\cite{Kao,dror,CH,Haldane}. This mechanism for 
cancelation of anomalies is discussed in detail in Appendix \ref{sec:CH}. 
 
%%%%%%%%%%%%%%%%%%%%%%%%%%%%%%%%%%%%%%%%%%%%%%%%%%%%%%%%%%%%%%%% 
\section{The saddle-point equation and the stripe solution} 
%%%%%%%%%%%%%%%%%%%%%%%%%%%%%%%%%%%%%%%%%%%%%%%%%%%%%%%%%%%%%%%% 
\label{saddlepoint} 
 
In the last section we constructed an effective action suitable for a stripe state. 
The effective action has two contributions: a ``bulk" piece and an ``edge" piece. In 
section \ref{sec:Model} we gave a simple, quadratic local expression for the ``bulk" 
contribution, valid at the one-loop level and for short range interaction. As we 
discussed above this form of the effective action is an RPA expression and it does 
not include the conventional Fock correction to the electron self-energy. 
The ``edge" contribution is due to charge fluctuations on the ``strings" discussed 
in the last section. Notice that at this level the description is still  
static. 
   
In this section we will obtain the saddle point configuration, $\varphi_{sp}$,  
for a stripe state of a partially filled  
$N-{\rm th}$ Landau Level ({\it i.e.} for $N-1<\nu <N$).  
Here we will use the effective action discussed in last section to construct a 
solution using the following procedure. We will take the configuration  
$\varphi_{sp}$ to be a smooth periodic function of the coordinate $y$ perpendicular 
to the stripes (which we take to be along the $x$ direction), 
\be 
\varphi_{sp}(x,y)\equiv\varphi_{sp}(y)=\varphi_{sp}(y+\lambda), 
\ee 
where $\lambda$, is the period of the stripe. Furthermore we will assume that the 
system has an extent $L_y=n_s \lambda$ along the $y$ axis where $n_s$ is the number of 
stripes. In what follows we will work with a fixed number of periods $n_s$ and 
find the value of the period $\lambda$ that minimizes the total energy at fixed but 
large $n_s$.   
  
We will construct the stripe state as follows.  
First, as we discussed in the previous section, 
we will regard the stripe state as a set of bulk regions 
separated by strings, representing the edges, {\it i.\ e.\/} the set of points of the plane where 
$\varphi_{sp}$ crosses the chemical potential $\mu$.  
We will construct an extremal solutions $\varphi_{sp}$ which is a smooth,  
differentiable and periodic function with wavelength (or period) $\lambda$. 
As we showed in section \ref{sec:Model}, in the ``bulk" regions the electron gas is  
incompressible. Thus, in these regions, the effective action $S_{\rm eff}$ is well approximated by  
a local function of the field $\varphi_{sp}$. Consequently, inside these regions,  
$\varphi_{sp}$  is just the solution of a simple equation, the saddle-point equation. 
The solution can then be constructed locally and it will be subject  
to appropriate boundary (matching) conditions on the curves representing  
the enclosing edges. For short range interactions the saddle-point equation is 
a partial differential equation whose solutions are easily constructed.  
In the stripe state there are two types of bulk regions, with   
filling factors $\nu=N$ and $\nu=N+1$ respectively, separated from each 
other by curves (or strings), the internal edges. We will denote by $\nu_T$ the 
effective filling factor of the partially filled landau level $N$. Thus, 
$\nu_T$ denotes the fractional area of the sample occupied by regions with 
$\nu=N+1$. Hence, $\nu_T$ is fixed by the number of electrons and by the magnetic 
field and it will be held fixed as we determine the optimal solution by varying 
over the period $\lambda$.  
A qualitative picture of this solution is depicted in fig.\ \ref{figstripe}. 
In the rest of this section we present the main results of this analysis,  
relegating the details to Appendix \ref{sec:mean-field}. 
 
It is convenient to define the dimensionless coupling constant  
$g \equiv 2\pi g \omega_c\ell^2$. We will also  
rescale the lengths, including the range $a$ of the potential, 
by $ x \to x \ell$ and $t\to t/\omega_c$, where $\ell$ is the magnetic 
length and $\omega_c$ is the cyclotron frequency. With this choice of units 
the saddle-point equation (SPE) in the bulk regions is given by 
\be 
- \left(\gamma(\varphi)+\frac{a^2}{g}\right)\nabla^2\varphi+ 
\frac{\varphi}{g}=\gamma(\varphi) \omega_c 
\label{saddle-point} 
\ee 
Eq.\  \ref{saddle-point}, is invariant under global translations and 
rotations, as well as under local gauge transformations. Homogenous and 
inhomogeneous 
solutions of Eq.\  \ref{saddle-point} were constructed in Ref.\ 
\cite{BLA}. 
The homogeneous solution, with $\gamma(\varphi)=N$, is $\varphi_N=N 
g\omega_c$ and it represents a uniform quantum Hall fluid state with filling factor 
$\nu=N$. Inhomogeneous solutions of this equation are also permitted and have 
the form 
\be 
\varphi=\varphi_N+\eta, 
\ee 
Since $\gamma(\phi_N+\eta)=N$, $\eta$ is the solution of  
\be 
\nabla^2\eta-\xi_N^2\eta=0, 
\label{obvious} 
\ee 
and $\xi_N^2\equiv {\displaystyle{\frac{1}{g N+a^2}}}$.  
 
The solutions $\eta(y)$ of Eq.\ \ref{obvious} are simple real exponential functions with suitably chosen 
coefficients. The condition on the function $\gamma$ implies that in a given region $\eta$ should 
satisfy the bounds  
\be 
(N+\frac{1}{2})\omega_c-\mu-\varphi_N< \eta <(N+\frac{3}{2})\omega_c-\mu-\varphi_N . 
\label{condition} 
\ee 
For a stripe with period $\lambda$ and effective filling factor $\nu_T$, for each period there are two 
incompressible regions. Since the solution is periodic it is sufficient to consider the fundamental interval 
$0\leq y <\lambda$ and the two incompressible regions meet at $y=\nu_T \lambda$. What matters here is that 
a smooth dependence of charge distribution on $y$ requires that 
the solution $\varphi(y)$ should be not only continuous at $y=\nu_T \lambda$ but also differentiable. Otherwise 
the charge distribution will not be differentiable and the energy of the state is necessarily 
larger. The solution thus constructed is then extended periodically beyond the fundamental period 
$[0,\lambda)$. Notice that in this construction the value of the chemical potential $\mu$ is determined from 
the value of the full solution $\varphi(y)$ at $y=\nu_T \lambda$. In Appendix 
\ref{sec:mean-field} we give explicit expressions for the function $\eta(y)$.  
 
In order to determine the optimum period $\lambda$ we now need to  
minimize the energy. To do that we will consider a stripe state with  
a fixed and finite number of periods $n_s$, for a 
system with a finite width $L_y$ commensurate  
with the number of stripes, {\it i.\ e.\/} $L_y=n_s\lambda$.  
 
Next we compute the total energy which is the sum of the ``bulk" energy associated with the solution 
$\varphi(y)$, and the energy of the ``edges" in the partially filled Landau level.  
In Appendix \ref{sec:mean-field} we give details of the solution and of the calculation  
of its ground state energy. There we show that the bulk contribution to the 
energy, computed from $S_{\varphi}[\varphi_{sp}]$, is a monotonically 
increasing  function of $\lambda$ (at fixed $n_s$), as it is expected for the 
Hartree term of the ground state energy.  
In particular, for large $\lambda$ the bulk energy is to an excellent approximation 
linear in $\lambda$ .    
 The energy due to the charge fluctuations at the ``edges" is 
obtained by integrating out the fermions near the regions where the  
solution crosses the chemical potential. 
This energy depends parametrically on the local profile of the stripe  
which acts as the effective 
electrostatic potential that creates the edges and  
it is a monotonically decreasing function of $\lambda$, which 
diverges as $\lambda\rightarrow 0$, and approaches zero 
as $\lambda\rightarrow \infty$.  
 
Next minimize the {\em total energy per period}.  
Since the bulk to the energy {\sl per period } 
is a monotonically increasing function of the period $\lambda$ 
(roughly linear), and the edge contribution {\sl per period} is a monotonically decreasing function of 
$\lambda$, there exists a finite value of the period 
$\lambda$ which minimizes the total energy per period. 
The result is a rather complicated function of the  
coupling constant $g$, of the  
Landau level index $N$ and of the effective filling factor $\nu_T$ of the  
partially filled Landau level. The 
solution simplifies considerably if the Landau level is large, $N \gg 1$, 
and  for $\nu_T=1/2$.  
In this limit we 
find that $\bar \lambda$, the optimal value of the period, is given by  
\begin{equation} 
{\bar \lambda}=4 \ell \; \sqrt{gN+a^2} \; \sinh^{- 
1}\left(\sqrt{\frac{g}{g+1}}\right)  
\label{eq:period} 
\end{equation} 
where $\ell$ is the magnetic length.  
For finite $N$, 
even for $N$ as low as $2$, the large $N$ expression turns out to be  
a good approximation. Notice that for reasonable values of the dimensionless  
coupling constant $g<1$, $\lambda\approx 3 \sqrt{N}\ell$. These results 
are in qualitative agreement with the  
more precise Hartree-Fock calculations \cite{fogler,chalker}.   
 
We also find that the solution changes very smoothly as a function of $\nu_T$ the  
vicinity of $\nu_T=1/2$. Thus, from now on we will restrict our 
discussion to the much simpler case of  
$N\gg 1$ and $\nu_T=1/2$.  In 
practice, there are few other details of  
the solution that we will need for the  
rest of the discussion. In 
particular in the following section we will use the solution explicitly to  
determine the velocity of the 
effective edge modes as well as to compute the elastic constants of the  
smectic phase (see Appendix 
\ref{sec:constants}). 
 
Thus, this procedure yields a finite value of the  optimal 
period $\lambda$. Once 
$\lambda$ is determined , we take the thermodynamic limit by just  letting $n_s \to \infty$. 
Notice that, in this process,  we actually vary the area of the system at fixed  
filling factor and fixed number of periods. In this process the number of 
particles is not necessarily kept fixed. Also, in this calculation,  
the chemical potential is not fixed either, as it is depends on the position of the stripes, and it is   
determined from the actual saddle-point solution that minimizes the energy. 
 
The variational approach that we followed here differs in a  
number of ways from the conventional 
Hartree-Fock approximation. In Hartree-Fock one works with a system with {\sl fixed 
size} in the thermodynamic limit, and looks for an extremum of the (free) energy 
density. In this approach there is a 
competition between the Hartree contribution, which favors stripes with  
$\lambda \to 0$, and the Fock term which favors stripes $\lambda \to \infty$, 
resulting in a state with finite period. In our construction we also found a 
Hartree term which favors a state with $\lambda \to 0$ but here the state is 
stabilized by the contribution from the edges (see Appendix \ref{sec:mean-field}). 
In the approach that we followed here, the energy associated with the edge fluctuations (usually ignored in 
Hartree-Fock) counter the preference of the Hartree term for stripes with vanishingly small period, by 
supplying a ``pressure" term which stabilizes the state. Notice, however, that if 
instead of minimizing the energy {\sl per period} we would have minimized the energy {\sl per unit 
total transverse length} $L_y$, we would not have found a minimum  since the edge contribution 
{\sl per unit length} is a constant, independent of the period $\lambda$. The resulting state 
would have had $\lambda=0$. Of course this is what happens in Hartree-Fock, in which case it is the Fock (or 
exchange) term of the {\sl energy density} what stabilizes the state. In any case, what will matter here is 
that it is possible to construct  a state with the correct properties, although a number of them, such 
as  the ground state energy are not as good as in Hartree-Fock. In any case most our results are fully 
consistent with the work of Refs. \cite{fogler} and \cite{chalker}  even at the quantitative  
level. The simpler approach that we used here   
has the advantage of being very intuitive and that it yields analytical results making the analysis of the 
fluctuations considerably simpler than in Hartree-Fock.  
 
%%%%%%%%%%%%%%%%%%%%%%%%%%%%%%%%%%%%%%%%%%%%%%%%%%%%%%%%%%%%%%%%%%% 
\section{Quantum Fluctuations and smectic symmetry} 
%%%%%%%%%%%%%%%%%%%%%%%%%%%%%%%%%%%%%%%%%%%%%%%%%%%%%%%%%%%%%%%%%%% 
\label{fluctuations} 
 
In this section we consider the effect of quantum fluctuations about the 
mean-field state found in the previous section. The low energy modes of the system  
in this state are smooth deformations of the location of the stripes on length scales 
long compared with the period of the stripe. These are the Goldstone modes of  
the broken translational symmetry. In terms of the Hubbard-Stratonovich field, these fluctuations are not 
small and cannot be treated simply as Gaussian perturbations since they do not have a restoring  
force. These fluctuations are similar to the zero modes of soliton systems and must be  
quantized exactly. On the other hand, small fluctuations of $\varphi$ are gapped, and, among other effects, 
they describe deformations of the stripes with a typical length scale shorter than the period $\lambda$. 
Ultimately, the main effects of these fluctuations is to renormalize the parameters of the low energy 
theory, including the forward scattering interaction between stripes. 
 
We will parametrize the low energy modes with a collective coordinate 
$u(x,y,t)$, that varies on long length scales $|x|,|y|>>\lambda$, and long times, 
$|t|>>1/\omega_c$, where $\omega_c$ is the cyclotron frequency. 
In this way, the set of functions that represent the low energy modes are 
given by smooth deformations of the saddle-point solution, 
\be 
\varphi=\varphi_{\rm sp}\left( y \alpha[u]+ u(x,y,t)\right) 
\label{deformedsp} 
\ee 
where $\alpha[u]$ is given by 
\be 
\alpha[u]=1-\frac{1}{2}\left(\frac{\partial u}{\partial x}\right)^2 
\ee 
is a small dilation of the $y$ coordinate needed to keep the period of the stripe constant,  
even for ``small" rotations $u\propto x$. This parametrization is sufficient to construct the (linearized) 
effective theory of the Goldstone modes, which has the form of a quantized elastic theory.  
 
In this parametrization, the $y$-coordinate of the $n$-th stripe is  
\be 
y=y_n \alpha[u]+u(x,y_n,t) 
\label{deformedstripes} 
\ee 
where $y_n$ is the coordinate of the $n$-th period of the stripe in the mean-field configuration. 
 
Therefore, we will split the Hubbard-Stratonovich field in two terms: a  
deformation of the saddle 
point parametrized by the displacement field $u$, and the high energy local  
fluctuations  
$\delta \varphi_u$: 
\be 
\varphi(x,y,t)= \varphi_{\rm sp}(y\alpha[u]+u)+ \delta\varphi_u 
\ee 
The effective action Eq. \ref{completeaction} now takes the form  
\bea 
\lefteqn{S_{\rm eff}[\varphi,u,\{\phi\}]=S(\varphi[u])+ 
S_{\phi}+S_I}\nonumber \\ 
&+& \frac{1}{2}\int d^3x d^3x' \delta\varphi_u(x)\frac{\delta^2 
S}{ \delta\varphi_u(x)\delta\varphi_u(x')}\delta\varphi_u(x)+\ldots \nonumber \\ 
&& 
\label{completeaction2} 
\eea 
In this equation $S(\varphi[u])$ is the action Eq.\ \ref{bulk} of the deformed  
saddle-point $\varphi[u]=\varphi_{sp}(y\alpha[u]+u)$,  
while $S_{\phi}$ and $S_I$ 
are given by Eq.\ \ref{Schiral0} and Eq.\ \ref{interaction} evaluated on the  
deformed stripes of Eq.\ \ref{deformedstripes}. In Eq.\ \ref{completeaction2} we  
have not taking into account couplings between $\delta\varphi$ and higher  
derivatives of $u$ since these interactions only give rise to irrelevant operators that only  
renormalize the coupling constants of the theory.   
 
Since the small fluctuations $\delta\varphi$ couple linearly with the charge  
density, they can be readily integrated out. Their net effect is a contribution to the forward  
scattering interaction among the edge modes. Thus, we get an effective action for  
the chiral edge fields of the form 
\be 
S_{\phi}=\int {\frac{d^2q d\omega}{(2\pi)^3}} \sum_{a,b} \frac{1}{2} 
\phi^a(\omega,{\vec q}) 
 \pi^{ab}(\omega,{\vec q})\phi^b(\omega,{\vec q})^*    
\label{fixedpoint} 
\ee 
where $a,b=\pm$ denotes chirality of the mode. The tensor $\pi^{ab}$ is given by  
\begin{widetext}
\be 
\pi^{ab}(\omega,{\vec q})= 
\left( 
\begin{array}{cc} 
-q_x\omega-q_x^2\left\{v+{\cal F}^{++}({\vec q})\right\}  & -q_x^2{\cal F}^{+- 
}({\vec q}) \\ 
& 
\\ 
-q_x^2{\cal F}^{-+}({\vec q}) &  q_x\omega-q_x^2\left\{v+{\cal F}^{-- 
}({\vec q})\right\} 
\end{array} 
\right) 
\label{Dalphabeta} 
\ee 
\end{widetext}
The effect of the integration over the high energy modes is encoded in the 
functions ${\cal F}^{ab}({\vec q})$. An explicit evaluation of these  
functions is given in Appendix \ref{sec:propagators}. 
 
Thus, we arrive to a long distance  effective action containing essentially two  
sets of  degrees of freedom: the long wavelength deformations  
$u(x,y,t)$, and the charge  density fluctuations $\phi^a(x,t)$. 
The effective action can be cast in the form,  
\be 
S=S_u+S_\phi+ S_I 
\label{utheta} 
\ee 
where the first term depends only on $u$, the second on the chiral fields  
$\phi_{\pm}$ and the third describes the interaction between charge and  
deformation  
through 
\be 
S_I=\sum_n\int dx~\partial_x\phi_n^a \varphi (y_n^a+u(x,y_n,t)) 
\label{varphitheta} 
\ee  
 
We see that we can obtain an effective theory for the stripe deformation  
integrating out  
the charge degrees of freedom. Conversely, integrating the deformation field we  
obtain a theory for the charge fluctuations. Of course these two actions  
contain the same physics.  
 
The form of $S_u$ is strongly constrained by symmetry. To begin with, $S_u$ it  
is a function only of the derivatives of $u$ since a constant shift in $u$ is just a global  
translation, which has no energy cost. In addition, a constant derivative along the direction of  
the stripe is equivalent to an infinitesimal global rotation, which in the absence of symmetry breaking 
fields is also an  exact symmetry  of the action. Thus, for small distortions, the effective action  
$S_u$ has the  form discussed in ref. \cite{BFKO} (see Eq.\ \ref{eq:Lsm}): 
\be 
S_u=-\int dxdydt  \left\{   \frac{Q}{2} 
\left(\frac{\partial^2 u}{\partial x^2}\right)^2 
+ \frac{\kappa_\perp}{2} \left(\frac{\partial u}{\partial  
y}\right)^2\right\}+\ldots 
\label{Ssmectic} 
\ee 
where $Q$ and $\kappa_\perp$ are elastic constants that will be given below.  
In momentum space $S_u$ becomes 
\be 
S_u=-\int \frac{d^2qd\omega}{(2\pi)^3}\;   
\left( \frac{Q}{2} q_x^4 + \frac{\kappa_\perp}{2} q_y^2+\ldots\right)|\tilde  
u(q)|^2 
\ee 
\noindent From the symmetry point of view, this action completely characterizes  
the smectic phase.  
In particular, it has the same symmetries of the free energy for a classical  
smectic\cite{deGennes}. 
Thus, the coefficient $Q$ is the compressibility of the system and the typical  
length scale 
$\xi=\sqrt{Q/\kappa_\perp}$ is the penetration length.  
 
However, the properties of the quantum smectic phase are not determined by  
symmetry alone since such arguments cannot determine 
the quantum dynamics of this phase. In order to find what is the dynamics of  the quantum  
smectic, we begin by noting that, at frequencies low compared with  $\omega_c$, the  
displacement fields $u$ {\sl do not} actually have a dynamics of their own. At this energy scale, 
their dynamics results solely from the coupling to the fluctuations of the gapless fields $\phi^a$,  
representing the ``edge modes" of each stripe. These are the only degrees of freedom with low energy  
states. Thus, the dynamics of the quantum smectic is controlled by the coupling between  
the displacement fields and the edge modes.  
 
In order to investigate the role of this coupling it is instructive to integrate out the degrees of 
freedom $\phi^a$, representing the edge modes, and to determine the form of the effective theory of the 
displacement fields $u$ alone. However, since the fields $\phi^a$ are gapless, the resulting effective 
dynamics of the displacement fields is non-local. Let us denote by $\varphi_n^a$ the deformed saddle-point 
solution for the $a$-th edge of the $n$-th stripe, {\it i.\ e.\/}  
\be 
\varphi_n^a\equiv \varphi_{sp}(y_n^a 
\alpha[u]+u(x,y_n,t)) 
\ee 
The effective action obtained upon integrating ou the edge modes has the form  
\be 
-\sum_{n,m}\int dxdx'dtdt' \frac{1}{2} \partial_x\varphi^a_n  
\left(\pi^{-1}\right)_{n,m}^{ab}\partial_{x'}\varphi^b_m 
\label{integration} 
\ee 
This expression has two contributions: static and dynamic.  
The static contribution is proportional to $\varphi^2$. This is just a renormalization of the coupling 
constant $g$  and at this level its only effect is to renormalize the coefficient $\kappa_\perp$ in  
Eq.\ \ref{Ssmectic}. Thus, in order to evaluate the dynamic contribution in what follows we will  
subtract the  static part from Eq.\ \ref{integration} and absorb it an a finite  
renormalization of $\kappa_\perp$ (see below).  
 
In order to write Eq.\ \ref{integration} in terms of the displacement field $u$  
we calculate explicitly the $x$-derivatives and write the resulting expression in the continuum limit  
on the $n$ variable. The form of the end result of this calculation 
is dictated by the symmetries of the classical smectic as well as by how the  
ground state stripe configuration transforms under spatial reflections.  
For the case that we have worked out in detail in 
this paper the stripe, {\it i.\ e.\/} the charge profile, is invariant under  
reflections about the middle of a stripe. This is a parity even state.   
A consequence of this symmetry is that the effective velocities of right  
and left moving edge fields on each stripe $\phi^a_n$ are the same.  
However, if the stripe state is not invariant under reflection, the parity odd  
piece of the solution forces the effective velocities of the left and right moving fields to be  
unequal, resulting in a different spectrum of low energy states. We will refer to this as to the parity  
odd state. Physically, a simple way to get an asymmetric state is to 
apply an in-plane electric field perpendicular to the stripe state (naturally, in a system with the 
center of mass pinned by the confining potential). This situation would yield ``unbalanced'' chiral 
excitations with different  velocities and couplings for the right and left movers. It is simple to show 
that the breaking of parity changes the spectrum from a $\omega \sim q_x^3$ dispersion to an $\omega \sim 
q_x^5$ law (at $q_y=0$ ). We will not discuss this case in detail. 
 
For a parity even stripe, the velocity of the chiral modes are equal,  
$v=v_R=v_L$ and it is a simple task to compute the dynamical term of the displacement field $u$. 
Putting all terms (both static and dynamic) together we find that the effective Lagrangian (in 
Fourier space) for the displacement fields $u$ is given by 
\bea 
{\cal L}[u]&=&\left[ 
\frac{{\bar v}^2}{2\kappa_\parallel \lambda^2 \ell^4} 
\left( 
\frac{-\omega^2}{\omega^2-\bar v^2 q_x^2} 
\right) 
-\frac{Q }{2}q_x^4-\frac{\kappa_\perp}{2} q_y^2 
\right]|\tilde u_q|^2 
\nonumber \\ 
&& \label{Susmectic} 
\eea 
where $\tilde u_q\equiv \tilde u(\omega,q)$,  
$\bar v$ is the renormalized velocity, $\lambda=4 g \sqrt{N} \ell$ is the period of 
the stripe in the limits $Ng+a^2>>g$, $g<<1$. In this limit, the elastic constants take the values 
\bea 
v               &=& \frac{g}{2} N^{-1/2}\omega_c \ell 
\\ 
\kappa_\parallel &=&{\displaystyle{\frac{\pi^2}{16 g^3 }}} N^{-3/2}\omega_c  
\\ 
\kappa_\perp     &=&\frac{5}{4} g N^{5/2} \omega_c \ell^{-2} 
\\ 
Q               &=&\frac{3}{128 \pi} g N  \omega_c  
\label{eq:constants} 
\eea 
where we have used that for $g$ small $\bar v \approx v$. 
A detailed derivation of these constants is given in Appendix \ref{sec:constants}. 
 
~From Eq.\ \ref{Susmectic} we immediately obtain the dispersion relation for  
the low energy excitations (in dimensionless form): 
\be 
\omega^2={\bar \kappa_\parallel}  q_x^2 
\left[\frac{Q q_x^4+ \kappa_\perp q_y^2} 
{1+ 
{\displaystyle{ 
\frac{{\bar \kappa_\parallel} }{v^2}\left( Q q_x^4+ \kappa_\perp q_y^2\right) 
}}} 
\right] 
\label{dispersion0} 
\ee 
where ${\bar \kappa_\parallel}= \kappa_\parallel \lambda^2 \ell^4$.  
Clearly, there is a crossover at the momentum scale  
$q^*=(v^2/Q{\bar \kappa_\parallel} )^{1/4}$ where the dispersion  
changes form cubic to a linear behavior:  
\bea 
\omega&=&\pm  
\sqrt{{\bar \kappa_\parallel}}  
|q_x|\sqrt{Q q_x^4+\kappa_\perp q_y^2},\;  
\mbox{for},\; q_x<<q^*   
\label{cubic}\\ 
\omega&=&\pm \bar v \;\; |q_x|,\; \mbox{for},\; q_x>>q^* 
\eea 
It is obvious from Eq. \ref{Susmectic} that the dynamics of the stripe  
deformations is non-local. This behavior is induced by the dynamics of the gapless ``internal  
edge states''.  
 
It is interesting to note that, in spite of this non-locality, it is  simple to  
find a local effective Landau-Ginsburg-like theory for the quantum unpinned smectic phase  
using two fields: (a) the Goldstone mode $u$, and (b)  an effective non-chiral scalar field $\phi$  
representing the charge fluctuation of each stripe. This is possible so because the $u$ field for a parity 
even stripe couples with  to a non-chiral linear combination of $\phi^+$ and $\phi^-$.  
The effective Landau-Ginsburg Lagrangian for the smectic phase, in dimensionful units, is given by  
\bea 
{\cal L}= && \frac{1}{\lambda \ell^2} \phi\partial_t u+  
\frac{\kappa_\parallel}{2v^2}(\partial_t \phi)^2- 
\frac{\kappa_\parallel}{2} (\partial_x \phi)^2 
 \nonumber  \\ 
&&-  \frac{Q}{2} 
\left(\partial_x^2 u\right)^2-\frac{\kappa_\perp}{2}\left(\partial_y  
u\right)^2  
\label{eq:fp} 
\eea 
The elastic constants of this effective action are given in Eq.\  
\ref{eq:constants}. 
 
Except for the second term, and up to a choice of units, the effective  
Lagrangian of eq.\ \ref{eq:fp} has exactly the 
same form as  the Lagrangian of Eq.\ \ref{eq:Lsm} which was introduced on 
phenomenological grounds in ref.\ \cite{BFKO}. The  term  
proportional to $(\partial_t 
\phi)^2$ is an irrelevant operator at the quantum Hall smectic fixed point.  
This irrelevant operator is 
responsible for the crossover discussed above at momenta $q_x \gg q^*$. In this  
regime  
the quantum smectic 
behaves effectively in the same way as a system of pinned stripes, the smectic  
metal phase of ref.\cite{EFKL}. 
 
 Using the values of the elastic constants of Eq.\ \ref{eq:constants} 
it is straightforward to estimate the scale at  
which the crossover between the cubic and linear dispersion takes place.  
Using the value of the stripe period known from Hartree-Fock  
calculations\cite{fogler}, to estimate a value for the coupling constant $g$,  
and our analytical expressions in the large $N$ limit, we find a crossover 
momentum scale at 
\be 
q^*\approx \frac{2}{\sqrt{\pi}} \frac{\ell^{-1}}{N^{3/8}}  
\label{cross} 
\ee  
where $\ell$ is the magnetic length. 
For instance, for $N=2$ $\ell q^*\approx 0.86$  
and  for $N=4$, $\ell q^*\approx 0.66$. This means that for lower Landau levels, 
the crossover takes place  at scales near the ultraviolet cutoff $1/\ell$.  
In this regime the dispersion relation, for an excitation with $q_y=0$, is  
cubic,  
\be 
\omega\approx 0.2 \;N^{1/2}\; \omega_c \left(\ell q_x\right)^3 
\ee 
where $\omega_c$ is the cyclotron frequency. For $N=2$ and for $q_x\ell=1/2$  
the typical frequency is $\omega\approx 4\times 10^{-2} \omega_c$. 
Thus, a light scattering experiment probing the system at wavevectors in the 
regime $\ell q_x\approx 0.5$, should be able to reach the  
cubic regime of the dispersion relation, which is the signature of the 
(unpinned) quantum Hall smectic phase. These wavevector correspond to a  
wavelength 
of the order of $2$ to $3$ times the stripe period, {\it i.\ e.\/} of the order  
of 
$1\mu m$. 
In higher Landau levels this effect should be more difficult to detect   
since the crossover scale decreases rapidly as $N$ increase 
(see Eq.\ \ref{cross}). 
 
In summary, we found an effective long wavelength description of  
the quantum smectic in full 
agreement with the general picture of stripe states as {\sl Electronic 
Liquid Crystal Phases}, as discussed in references \cite{nature,FK}.  
 
%%%%%%%%%%%%%%%%%%%%%%%%%%%%%%%%%%%%%%%%%%%%%%%%%%%%%%%%%%%%%%% 
\section{Conclusions and Open Problems} 
%%%%%%%%%%%%%%%%%%%%%%%%%%%%%%%%%%%%%%%%%%%%%%%%%%%%%%%%%%%%%%%%%%% 
\label{conclusions} 
 
In this paper we have derived the effective theory for the low energy degrees of 
freedom of the quantum Hall smectic phase introduced phenomenologically in ref.\  
\cite{BFKO}. 
  
The quantum Hall smectic phase of the 2DEG breaks spontaneously both translation 
invariance (in the direction perpendicular to the stripes) and rotational 
invariance. In our approach 
the quantum Hall smectic is pictured as a pattern of locally  
incompressible regions separated by 
dynamical edges  
determined by an effective (Hubbard-Stratonovich) dynamical potential. The 
resulting ``edge modes" are thus coupled to the dynamical fluctuations of 
the shape of the incompressible regions, represented by a set of displacement  
fields.  Consequently, the low energy physics 
of the quantum Hall smectic is described in terms of two coupled and canonically conjugate
fields: (a) the displacement fields $u$ that describe the geometry 
of the quantum Hall smectic, and (b) the charge degrees of freedom represented by the  
non-chiral ``Luttinger field'' $\phi$.  
 
While the form of the static part of the action is dictated entirely by symmetry, in general the dynamics 
depends on the particular details of the model. However, in this case, the effective dynamics 
is dictated by the Lorentz force which governs the motion of charged particles in electromagnetic fields.
An important consequence is that it requires that the charge and displacement  fields to become
canonical conjugate to each other. 
In reference \cite{BFKO} we discussed the consequences of this effective theory as a fixed point.
There it is also discussed at length the issue of the stability of the quantum Hall smectic phase and
a possible transition to a 
crystalline state, both of which remain still interesting and open problems.

In this paper we used the effective low energy theory to determine the  spectrum of collective  
modes of the quantum Hall smectic and found that they obey an  $\omega \propto q_x^3$ law. It should be 
possible to detect these modes in Raman scattering experiments. 
Furthermore, we also found that there exists an energy and momentum scales, determined by  
irrelevant operators,  above which  the quantum Hall smectic behaves like a two-dimensional array of
Luttinger liquids, a smectic metallic state. 
 
An open and very interesting question is the possible existence of a  
quantum nematic state of the 2DEG in large magnetic fields at zero temperature. 
This is an important question both conceptually and experimentally as it appears  
to be consistent with the experimental data\cite{FK,FKMN}. Recently, in ref.\ 
\cite{vadim} a theory of a quantum  nematic Fermi fluid at zero external magnetic field as an 
instability of a Fermi  liquid state was presented. It will particularly interesting to construct a 
theory of the {\sl quantum} melting of the quantum Hall smectic by a 
dislocation-antidislocation unbinding mechanism. A key ingredient of such a  theory is  
the quantum mechanical origin of charge quantization in the smectic.  
Work along these lines is currently in progress. recently, Wexler and Dorsey\cite{nematic} used
Hartree-Fock calculations to determine the effective elastic constants of a quantum Hall nematic phase and
used them to estimate the critical temperature of the quantum Hall nematic-isotropic transition.
  
Finally, another open question of interest is the  possible 
transition to a paired quantum Hall state\cite{read-moore,gww}. Our stability  analysis shows 
that it is possible to have a direct phase transition to a paired state.  This possibility is 
supported by exact diagonalization studies in small systems.\cite{HR,rezayi} However it is quite 
likely that there is a complex phase diagram, such as the one discussed in ref.\ \cite{FK}, 
including a nematic phase, various crystalline phases and incompressible fluid phases such as 
the paired quantum Hall state. 
 
\section{Acknowledgments} 

We are profoundly indebted to S.\ Kivelson and V.\ Oganesyan who helped us with numerous 
and pointed questions and discussions throughout this work. 
We thank  H.\ Fertig, M.\ P.\ A.\ Fisher, B.\ I.\ Halperin, T.\ C.\ Lubensky and  A.\ H. 
MacDonald for useful discussions. 
This work was supported in part by grants of the National Science 
Foundation numbers DMR98-08685 (SAK) and DMR98-17941 (EF). D.\ G.\ B. was  
partially supported by  the University of the State of Rio de Janeiro, Brazil  
and  by the Brazilian agency CNPq through a postdoctoral fellowship. 
 
%%%%%%%%%%%%%%%%%%%%%%%%%%%%%%%%%%%%%%%%%%%%%%%%%%%%%%%%%%%%%%%%%%% 
\appendix 
%%%%%%%%%%%%%%%%%%%%%%%%%%%%%%%%%%%%%%%%%%%%%%%%%%%%%%%%%%%%%%%%%%% 
 
%%%%%%%%%%%%%%%%%%%%%%%%%%%%%%%%%%%%%%%%%%%%%%%%%%%%%%%%%%%%%%%%%%% 
\section{Mean-Field Theory of the Stripe State} 
\label{sec:mean-field} 
%%%%%%%%%%%%%%%%%%%%%%%%%%%%%%%%%%%%%%%%%%%%%%%%%%%%%%%%%%%%%%%%%%% 
In this appendix we give the details of the construction of the  
saddle-point solution  
for the stripe state. 
As discussed in Section \ref{saddlepoint} we must first construct  
solutions of Eq.\ \ref{Svarphi}. 
In particular we will 
seek solutions which within a period $\lambda$ have the form 
\be 
\varphi(y)=\left\{ 
\begin{array}{lcc} 
\varphi_N(y) &\mbox{for}& 0<y<\nu\lambda \\ 
\varphi_{N+1}(y) &\mbox{for}& \nu\lambda<y<\lambda 
\end{array} 
\right. 
\label{stripesolution} 
\ee 
 where $\varphi_N$ and $\varphi_{N+1}$ are general solutions of the Eq.\ 
\ref{saddle-point}, with $\gamma=N$ and $\gamma=N+1$ respectively. In 
other terms, in the region of the plane where $\gamma=N$ all Landau 
levels up to and including the level $N-1$ are completely filled. (Here, 
for the sake of simplicity we are ignoring spin.) The filling factor 
$\nu$, with 
$0<\nu<1$, is the effective filling factor of the partially filled 
$N$-th Landau level. A general solution of this type reads 
\be 
\varphi_N(y)=g\omega_c N+ a_+ e^{\xi_N y}+ a_-e^{-\xi_N y} 
\label{sol} 
\ee 
Smooth periodic functions satisfy the boundary conditions: 
\bea 
\varphi_N(0)&=&\varphi_{N+1}(\lambda) \nonumber \\ 
\varphi'_N(0)&=&\varphi'_{N+1}(\lambda) \nonumber \\ 
\varphi_N(\nu\lambda)&=&\varphi_{N+1}(\nu\lambda) \nonumber \\ 
\varphi'_N(\nu\lambda)&=&\varphi'_{N+1}(\nu\lambda) 
\eea 
These conditions determine completely the coefficients $a_+$ and $a_-$ of 
Eq.\  \ref{sol}. The explicit solution is: 
\bea 
\lefteqn{\varphi_N(y)=g\omega_c\left\{ \right. 
N+ } 
\nonumber \\ 
&& \frac{ \xi_{N+1}\sinh\left\{\frac{\lambda}{2}[\xi_{N+1}(1-\nu)]\right\}} 
{\sigma(\xi,\nu)} 
 \cosh\left[\xi_N(y-\frac{\nu\lambda}{2}) \right] 
 \left .\right\} \nonumber \\ 
 && 
\label{phiN} 
\eea 
\bea 
\lefteqn{\varphi_{N+1}(y)=g\omega_c\left\{ \right. 
N+1-} 
\nonumber \\ 
&& 
\frac{\xi_N\sinh\left\{\frac{\lambda}{2}[\xi_{N}\nu]\right\}} 
{\sigma(\xi,\nu)} 
\;\;\cosh\left[\xi_{N+1}(y-\frac{\lambda}{2}(1+\nu)) \right] 
\left. \right\}  
\nonumber \\ 
&& 
\label{phiN+1} 
\eea 
where  
\bea 
\sigma(\xi,\nu)&=& 
\xi_{N+1}\cosh\left\{\xi_N\frac{\nu\lambda}{2}\right\} 
\sinh\left\{\xi_{N+1}\left(\frac{1-\nu}{2}\right)\lambda\right\}+ 
\nonumber \\ 
&& 
\xi_{N}\sinh\left\{\xi_N\frac{\nu\lambda}{2}\right\} 
\cosh\left\{\xi_{N+1}\left(\frac{1-\nu}{2}\right)\lambda\right\} 
\nonumber \\ 
&& 
\eea 
 
Thus, we found a family of solutions parametrized by the wavelength $\lambda$. 
A qualitative picture of this solution is depicted in figure \ref{figstripe}. 
As we already explain in section \ref{saddle-point} we fix $\lambda$,  
by   minimizing the {\sl total} energy per period.  
 
%%%%%%%%%%%%%%%%%%%%%%%%%%%%%%% 
\subsection{The energy of the saddle-point configuration} 
%%%%%%%%%%%%%%%%%%%%%%%%%%%%%%% 
There are two contributions to the total energy: 1) the bulk energy, and 
2) the energy of the chiral edges. 
%%%%%%%%%%%%%%%%%%%%%%%%%%%%%%% 
\subsubsection{The bulk energy} 
 
\begin{figure}
\begin{center}
\leavevmode
%\vspace{.2cm} 
\noindent
%\hspace{1.0 in}
\epsfxsize=7 cm
\epsfysize=6.0 cm 
\epsfbox{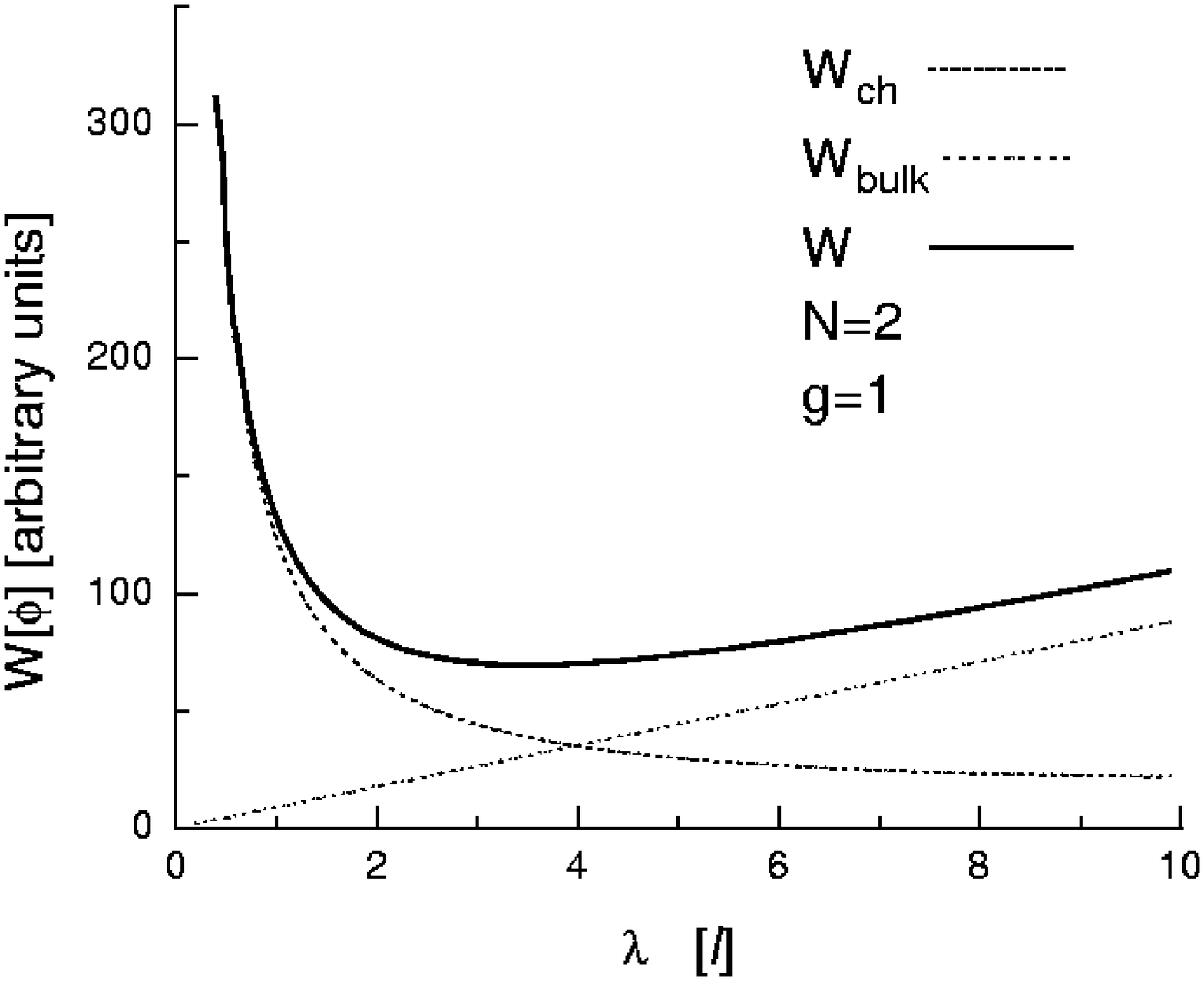} 
\end{center}
\caption{Energy of the saddle-point solution.} 
\label{fig2} 
\end{figure}

To calculate the energy of the stripe solution we simply replace Eq.\ 
\ref{stripesolution},  Eq.\ \ref{phiN}, and   Eq.\ \ref{phiN+1} into  Eq.\ 
\ref{Svarphi} to find 
\bea 
W_{\rm bulk}&=&\frac{1}{4\pi}\int d^2x~\gamma(\varphi) \left\{ 
\varphi+\gamma(\varphi) \omega_c 
\right\} \nonumber \\ 
&=&\frac{n_p  L_x}{4\pi}\int_0^\lambda dy~ 
\gamma(\varphi) \left\{ 
\varphi+\gamma(\varphi)\omega_c 
\right\} 
\eea 
where $n_p$ is the number of periods in the sample and $L_x$ is the 
length in the $x$ direction. The energy per period, per unit $x$-length has the  
form 
\be 
\tilde W_{\rm bulk}=\lambda \bar W(N)+W(\lambda) 
\label{bulkenergy} 
\ee 
where 
\be 
\bar W(N)={\frac{\omega_c}{2}} (1+g) \left[ 
N^2\nu+(N+1)^2 (1-\nu)\right] 
\ee 
and 
\bea 
W(\lambda)&=&g\omega_c \frac{1}{\xi_N} 
\left\{N\frac{\xi_{N+1}}{\xi_N}-(N+1)\frac{\xi_{N}}{\xi_{N+1}} \right\} 
 \nonumber \\ 
&\times& 
\frac{\sinh\left\{\frac{\lambda}{2}\xi_N\nu\right\} 
\sinh\left\{\frac{\lambda}{2}\xi_{N+1}(1-\nu)\right\}} 
{\sinh\left\{\frac{\lambda}{2}[\xi_N\nu+\xi_{N+1}(1-\nu)]\right\}} 
\eea 
Since $W(\lambda)$ is  bounded, ${\tilde W}_{\rm bulk}$ 
is a monotonically increasing function 
of $\lambda$ (with $W_{\rm bulk}(0)=0$). For large $\lambda$, 
the first term of Eq.\ \ref{bulkenergy} dominates, and in that limit ${\tilde  
W}_{\rm bulk}$ 
is essentially a linear function of $\lambda$, see fig. \ref{fig2}. 
%%%%%%%%%%%%%%%%%%%%%%%%%%%%%%%%%%%%%%%%%%%%%%%%%%%%%%%%%%%%%%%%%%%%%%%%%%% 
%%%%%%%%%%%%%%%%%%%%%%%%%%%%%%%%%%%%%%%%%%%%%%%%%%%%%%%%%%%%%%%%%%%%%%%%%%% 
%%%%%%%%%%%%%%%%%%%%%%%%%%%%%%% 
\subsubsection{The energy of the chiral edges.} 
 
As we already pointed out, the discontinuities of the function $\gamma(\varphi)$ 
determine a set of one-dimensional curves (``strings") where the chiral degrees  
of 
freedom reside. At the level of the saddle-point solution these are static 
straight lines at $y=n\lambda$ and $y=(n+\nu)\lambda$, with $n$ integer. 
 
Upon integrating over the array of chiral bosons, and taking 
into account that $\varphi$ is static, we can calculate the contribution of the 
edges to the energy per period. Since the saddle-point solution 
$\varphi$ is independent of $x$, 
and $\varphi(0)=\varphi(\nu\lambda)$ and $\partial_y\varphi(0)=- 
\partial_y\varphi(\nu\lambda)$, we find that the energy per period per unit  
length 
(along the $x$ axis) is 
\be 
\tilde W_{\rm edge}= 
\frac{\varphi(0)^2 }{\left|\partial_y 
\varphi(0)\right|} 
\ee 
In terms of the explicit solution Eq.\ \ref{stripesolution}, Eq.\ 
\ref{phiN}, and Eq.\ \ref{phiN+1}, this energy reads 
\bea 
\tilde W_{\rm edge}&=&\frac{g \omega_c }{\xi_{N+1}} 
\frac{\sinh\left(\frac{\lambda}{2}[\xi_N\nu+\xi_{N+1}(1-\nu)]\right)} 
{\sinh\left(\frac{\lambda}{2}\xi_{N+1}[1-\nu]\right) 
\sinh\left(\frac{\lambda}{2}\xi_{N}\nu\right)} 
\nonumber \\ 
\times 
\left[ 
N+\right. 
&& 
\left. 
\frac{\xi_{N+1}}{\xi_N} 
\frac{\sinh\left(\frac{\lambda}{2}\xi_{N+1}[1-\nu]\right) 
\cosh\left(\frac{\lambda}{2}\xi_{N}\nu\right)} 
{\sinh\left(\frac{\lambda}{2}[\xi_N\nu+\xi_{N+1}(1-\nu)]\right)} 
\right]^2 \nonumber \\ 
&&\label{edgeenergy} 
\eea 
For large $\lambda$ this function approximates exponentially fast a constant, 
and it diverges like $1/\lambda$ for small values of the period. 
 
Thus, the total energy per period of the saddle-point solution is 
\be 
\tilde W= \tilde W_{\rm bulk}+ \tilde W_{\rm edge} 
\label{totalenergy} 
\ee 
 where $\tilde W_{\rm bulk}$ and $\tilde W_{\rm edge}$ are given by Eq.\ 
\ref{bulkenergy} and Eq.\ \ref{edgeenergy} respectively. In figure \ref{fig2} 
we 
depict these functions. We see that the competition between bulk and edge 
energies  yields a stable and finite value of the 
period $\bar\lambda$. 
%%%%%%%%%%%%%%%%%%%%%%%%%%%%%%%%%%%%%%%%%%%%%%%%%% 
\subsection{The $N \gg 1$ limit} 
 
\begin{figure}
\begin{center}
\leavevmode
%\vspace{.2cm} 
\noindent
%\hspace{1.0 in}
\epsfxsize=7 cm
\epsfysize=7 cm
\epsfbox{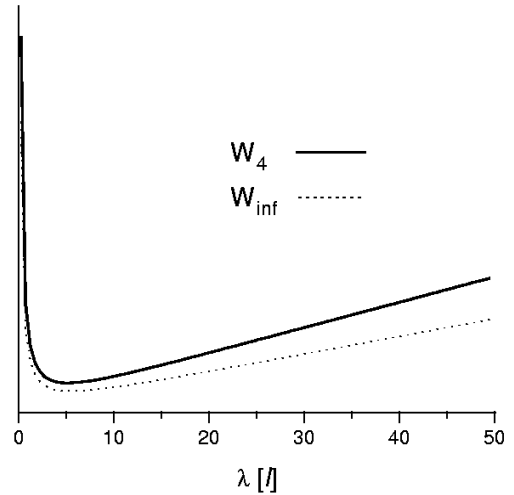} 
\end{center} 
\caption{Comparison between the energy for $N=4$ and the $N\to\infty$ limit} 
\label{fig3} 
\end{figure}

In order to clarify the dependence of the optimal period $\bar\lambda$ with the 
microscopic 
parameters of the theory it is convenient to consider particular limits where 
the 
expressions for the energy become tractable. 
In particular, it is possible to find an explicit analytic result for the 
period 
of the stripe in the limit $N \gg 1$ ({\sl c.\ f.\/} ref.\ 
\cite{chalker}). 
In this limit, $\xi_N/\xi_{N+1}=1+ {\cal O} (1/N) $, and the 
energy $\tilde W$ given by Eq.\ \ref{totalenergy} becomes 
\bea 
\tilde W=g \omega_c \frac{N^2}{\xi_N}&&\!\!\!\!\!\! 
\left\{ 
\left(\frac{1+g}{g}\right)\frac{\lambda \xi_N}{2} 
\right. \nonumber \\ 
&+& \left. 
\frac{\sinh\left(\frac{\lambda}{2}\xi_{N}\right)} 
{\sinh\left(\frac{\lambda}{2}\xi_{N}[1-\nu]\right) 
\sinh\left(\frac{\lambda}{2}\xi_{N}\nu\right)} 
\right\} 
\label{totalenergyN} 
\eea 
Note that $\lambda$ always appears in the combination $\lambda\xi_N/2$. 
This means that the natural scale for the period is $\xi_N/2$. 
In terms of the variable $\bar x=\frac{\lambda\xi_N}{2}$, the extremal 
condition 
\be 
\frac{d\tilde W(\bar x)}{d\bar x}=0 
\ee 
becomes 
\be 
\nu \; {\rm csech}^2(\nu\bar x)+(1-\nu) \; {\rm csech}^2((1-\nu)\bar 
x)=\frac{g+1}{g} 
\label{transcendental} 
\ee 
For $\nu=1/2$ we find the explicit solution 
\be 
\bar x=2 \sinh^{-1} \left\{ \sqrt{\frac{g}{g+1}}\right\} 
\ee 
Thus, the period of the stripe, for large $N$ and $\nu=1/2$, is 
\be 
\bar\lambda=4 \sqrt{gN+a^2} \sinh^{-1} \left\{ \sqrt{\frac{g}{g+1}}\right\} 
\; \ell 
\label{eq:period2} 
\ee 
Eq.\ \ref{eq:period2} implies that the period of the stripe is set by a 
combination of the cyclotron radius of the partially filled Landau 
level ${\sqrt{N}} \ell$, the range of the interaction $a$, and a 
function of the dimensionless coupling constant $g$. In particular, the 
wavelength of the stripe state is of the order of the cyclotron radius 
only in the limit in which the dimensionless range of the interaction is small, 
$a \ll {\sqrt{gN}}$. In this limit, the result of Eq.\ \ref{eq:period2} 
agrees with the estimates of Koulakov and coworkers\cite{fogler}. 
It turns out that expression Eq.\ \ref{eq:period2} is a very good 
approximation even for small values of $N$. In fig.\ \ref{fig3} we 
compare the numerical solution of the period for $N=4$ with the large $N$ 
approximation. 
Notice that the position of the optimal value of $\lambda$ is 
essentially the same for both curves. 
%%%%%%%%%%%%%%%%%%%%%%%%%%%%%%%%%%%%%%%%%%%%%%%%%%%%%%%%%%%%%%%%%%%%%%%%%%% 
%%%%%%%%%%%%%%%%%%%%%%%%%%%%%%%%%%%%%%%%%%%%%%%%%%%%%%%%%%%%%%%%%%%%%%%%%%% 
 
We can also solve equation (\ref{transcendental}) when $\nu\approx 1$. In this 
case the period of the stripe is written as 
\be 
\bar\lambda=\left(\frac{2}{1-\nu}\right)\sqrt{\frac{g}{g+1}}\sqrt{gN+a^2}\ell 
\label{eq:period1} 
\ee 
It turns out that the period increases as the filling factor 
increases (away from $1/2$). In any case, we expect that 
for  filling factors not too close to $1/2$ 
the stripe state should become unstable 
to other types of phases, such as a  bubble phase\cite{fogler}, 
a stripe crystal\cite{FK,MF,fertig} 
and  possibly a quantum nematic phase\cite{FK,FKMN}.

%%%%%%%%%%%%%%%%%%%%%%%%%%%%%%%%%%%%%%%%%%%%%%%%%%% 
\section{The geometrical couplings} 
\label{sec:geometry} 
%%%%%%%%%%%%%%%%%%%%%%%%%%%%%%%%%%%%%%%%%%%%%%%%%%% 

In this section we  derive the effective action for the chiral edge states of 
the 
stripe state. Qualitatively this problem is very much analogous to that of the 
fermion zero modes bound to a dynamical domain wall\cite{fosco}. Here we will 
show how a geometrical electric field is induced on the stripe due to its 
dynamics. 
 
The action for two-dimensional fermions in a magnetic field and a time 
dependent background potential $\varphi$ is 
\be 
S=\int d^3z~ \left[\psi^*(z)[iD_0-\mu-e \varphi]\psi(z) 
+\frac{1}{2m} 
\left|{\bf D}\psi(z)\right|^2\right] 
\label{action} 
\ee 
where ${\bf D}_i=\partial_i+i A_i$ and $\vec\nabla\times {\vec A}=B {\hat z}$. 
 
The Hubbard-Stratonovich field $\varphi$ behaves like a scalar potential 
in electrodynamics. As such it (adiabatically) deforms of the Landau levels. If 
the 
topmost filled Landau level $N$ crosses the chemical potential at some set of 
smooth curves, 
\be 
\varphi\left( 
\bar x(s,t),\bar y(s,t), t 
\right)=E_N-\mu 
\label{surface} 
\ee 
then the system has gapless excitations with support on these curves, which 
thus behave as dynamical edges. Here $E_N$ is the energy of the $N^{\rm th}$ 
Landau level. 
This equation 
defines a time dependent string-like object, a two-dimensional surface embedded 
in a 
Euclidean three dimensional space-time, whose position is defined by $\vec 
R(s,t)=(\bar x(s,t),\bar y(s,t))$. For the most part it will be 
sufficient to consider only one edge at a time. 
 
Let us expand the potential $\varphi$ around its constant value on the string 
\be 
\varphi(x,y,t)= 
\varphi(\bar x,\bar y,t)+ \tilde v \hat n_\mu x_\mu 
\ee 
where 
\be 
\tilde v=\left|\partial_\mu\varphi\right|= 
\sqrt{\left(\frac{\partial\varphi}{\partial t}\right)^2 
+\left|\vec\nabla\varphi \right|^2} 
\ee 
and $n_\mu$ is a unit vector perpendicular to the surface defined by 
Eq.\ \ref{surface}. 
The effective action in this  approximation is 
\be 
S=\int d^3z~ \left[\psi^*\left(iD_0-E_N+\tilde v \hat n_\mu x_\mu\right)\psi 
+\frac{1}{2m} 
\left|{\bf D}\psi\right|^2\right] 
\label{actionapp} 
\ee 
To proceed with the calculation, we rewrite the action in generalized 
coordinates 
\[ 
x_\mu=(x_0,x_1,x_2) \rightarrow  \xi_\mu=(\xi_0,\xi_1,\xi_2) 
\] 
The coordinate transformation and the metric is defined by 
$\xi_\mu=\frac{\partial\xi_\mu}{\partial x_\nu}~x_\nu$ and 
$g_{\mu\nu}=\frac{\partial x^\alpha}{\partial\xi^\mu} \frac{\partial 
x^\beta}{\partial\xi^\nu} \delta_{\alpha\beta}$ respectively. 
It is convenient to choose a coordinate system defined by 
\be 
\xi_2=\hat n_\mu x_\mu 
\ee 
in such a way that the differential quadratic form is given by 
\be 
ds^2=d\xi_2^2+g_{ab}d\xi^ad\xi^b~~~~{\rm where}~~~~a,b=0,1 
\ee 
In these coordinates the domain wall is defined by the equation $\xi_2=0$ 
and the action of Eq.\ \ref{actionapp} now reads 
\bea 
S&=&\int d^3\xi~\sqrt{g} 
\left[\psi^*(\xi)[i\left(\frac{\partial\xi_\mu}{\partial t}\right) 
\frac{\partial~}{\partial\xi_\mu}+\tilde v \xi_2-E_N]\psi(z)\right. 
\nonumber \\ 
&+&\left.\frac{1}{2m}\eta^{\mu\nu} 
\left|{\bf D}\psi(\xi)\right|^2\right] 
\eea 
where we have used the notation 
\be 
\eta_{\mu\nu}=\frac{\partial\xi_\mu}{\partial x_i} 
\frac{\partial\xi_\nu}{\partial x_i}\; , \qquad 
g= |{\rm det} g_{\mu \nu}| 
\ee 
It is now convenient to choose the Landau gauge in the new coordinates 
\be 
A_0=0\mbox{~~~~,~~~~}A_1=B\xi_2 
\mbox{~~~~,~~~~}A_2=0 
\ee 
Next we expand the Fermi field $\psi$ in the new coordinates and find 
\be 
\psi(\xi_0,\xi_1,\xi_2)=\sum_{n,p}C_{n,p} X^n(\xi_1)Y^n_p(\xi_2)T^n(\xi_0) 
\label{expansion} 
\ee 
The set of functions $\{X^n(\xi_1), Y^n_p(\xi_2), T^n(\xi_0)\}$ constitute 
complete 
basis, and 
satisfy the eigenvalue equations 
\bea 
i\partial_0 T^n&=&\omega_n T^n \\ 
i\partial_1 X^n&=&k_n X^n 
\eea 
\bea 
\lefteqn{\left\{ 
-\frac{\eta^{\mu 2}\eta^{\mu 2}}{2m}\partial_2^2- 
2B\eta^{\mu 1}\eta^{\mu 2} 
\left(\xi_2-\frac{m \tilde v}{(\eta^{\mu 1})^2}B^2\right) 
\partial_2 \right. } \nonumber \\ 
&+&\left.\frac{(\eta^{\mu 1})^2}{2m} \xi_2^2 
-E_N\right\} 
Y^n_p=\lambda_{n,p} Y^n_p 
\label{osc} 
\eea 
Eq.\ \ref{osc} is nothing but the eigenvalue equation for the linear harmonic 
oscillator 
(in generalized coordinates). By inspection we see that  Eq.\ \ref{osc}  has a 
zero mode 
$\lambda_{n,P}=0$ provided $E_N$ is the energy of a Landau level in the 
undistorted 
coordinates. Upon substitution Eq.\ \ref{expansion} into Eq.\ \ref{actionapp}, 
using the usual orthogonality relations for the oscillator eigenfunctions, and 
after factoring out the zero mode from the rest of the spectrum, 
we find that the effective action for the zero mode is given by 
\be 
S_0=\int d^2\xi~\sqrt{g}~\frac{1}{2}\left\{ 
\psi_0^*(\xi){\cal D}\psi_0(\xi)- 
{\cal D}\psi_0^*(\xi)\psi_0(\xi)\right\} 
\ee 
where 
\be 
{\cal D}= 
\left(\frac{\partial\xi_0}{\partial t}-v \eta^{\mu 1}\eta^{\mu 0}\right) 
\frac{\partial~}{\partial \xi_0}- 
\left(v \eta^{\mu 1}\eta^{\mu 1}-\frac{\partial\xi_1}{\partial t}\right) 
\frac{\partial~}{\partial \xi_1} 
\label{generalchiral} 
\ee 
Here we have defined $v \equiv {\tilde v}/B$. 
 
For the problem of interest here, we will specialize these results 
to the case of a stripe whose mean position is a straight line along the $x$ 
axis, 
as defined in the saddle-point approximation of section \ref{saddlepoint}. 
To this aim in mind, we define a coordinate system as follows 
\bea 
x_0&=&\xi_0 \nonumber \\ 
x_1&=&\xi_1 \nonumber \\ 
x_2&=& \delta(\xi_0,\xi_1)\equiv 
\frac{\delta\varphi(\xi_0,\xi_1,x_2=0)}{\left|\partial_{x_2}\varphi(x_2=0)\right 
|} 
\eea 
where $\delta(\xi_0,\xi_1)$ is an infinitesimal local displacement of the 
position 
of the edge. In this 
coordinate system the effective action can be cast in the form 
\be 
S=\int d^2\xi~\sqrt{g}~\left\{ 
\psi_0^*(\xi)\left(\partial_0-v \partial_1\right)\psi_0(\xi)+ 
\Gamma\psi_0^*(\xi)\psi_0(\xi)\right\} 
\label{chiral} 
\ee 
 where $\Gamma=\frac{1}{2}g^{-\frac{1}{2}}{\cal D}g^{\frac{1}{2}}$. 
This quantity couples in the same way that a gauge field couples to a chiral 
zero mode. Notice that this gauge field  looks like a pure gauge and as 
such it would seem that it should have no effect on the theory. That 
would be indeed the case if the theory of the zero modes was gauge invariant 
all on its own right. However, this is not the case here as these modes 
have a gauge anomaly. 
 
Eq.\ \ref{chiral} is precisely the action of a one dimensional chiral fermion 
in 
curved spacetime. It is very well known that this system is anomalous and as 
a consequence the divergence of the current is proportional to the curvature 
of the space-time. As a matter of fact, the current $J$ of the 
chiral fermion satisfies, 
\be 
\partial_0 J_0+v\partial_1 J_1={\rm det}\left(\frac{\partial^2 
\delta(\xi_0,\xi_1)}{\partial_{\xi_a}\partial_{\xi_b}}\right) 
\label{div} 
\ee 
to leading order in $\delta(\xi_0,\xi_1)$. 
Therefore, within this approximation, Eq.\ \ref{div} reduces to the  divergence 
of the edge current in Cartesian coordinates. Hence, we are led to interpret 
the quantity ${\rm det}\left(\frac{\partial^2 
\delta(\xi_0,\xi_1)}{\partial_{\xi_a}\partial_{\xi_b}}\right)$ as an {\sl 
induced 
geometrical electric field} given by 
\be 
E_{\rm 
geom}=\frac{1}{v}\partial^2_1\delta\partial_0^2\delta- 
(\partial_0\partial_1\delta)^2 
\ee 
The same expression for the  geometrical electric field was also derived 
for a system of Dirac fermions with time-dependent domain walls\cite{fosco}. 
It is straightforward to see that $E_{\rm geom}$ is generated by the  
``dynamical electromagnetic potential'' 
\bea 
A^{\rm dyn}_0&=& \partial_1 \delta \partial^2_0\delta  \\ 
A^{\rm dyn}_1&=& \partial_1\delta \partial_1\partial_0\delta 
\eea 
We have used this type of couplings on the second line of Eq.\  
\ref{interaction}. 
  
%%%%%%%%%%%%%%%%%%%%%%%%%%%% 
\section{Charge conservation, cancelation of anomalies and the Callan-Harvey 
effect.} 
%%%%%%%%%%%%%%%%%%%%%%%%%%%% 
\label{sec:CH} 
 
In this Appendix we show that charge conservation in a stripe state 
is realized thorough an anomaly cancelation mechanism that includes 
the effects of dynamical edges. 
 
The problem that needs to be addressed here is that we have separated the 
dynamical 
degrees of freedom into a ``bulk'' piece , given by Eq.\ \ref{Sa} and Eq.\ 
\ref{Svarphi}, 
and an ``edge'' piece, Eq.\ \ref{Schiral0}. It turns out that the $U(1)$ gauge 
transformation of the external gauge field is not a symmetry of each part of 
this action 
separately, but instead it is a symmetry of the full system. this 
problem is quite familiar in the physics of the 
QHE\cite{wen,frohlich,Haldane,Kao}. The main difference in the problem 
of interest here is that the edges are not static. 
This subtle cancelation of anomalies is an example of the well known 
Callan-Harvey mechanism\cite{CH}. 
 
To illustrate the point, let us consider 
a general time dependent gauge transformation, 
\be 
{\tilde A}_0\rightarrow {\tilde A}_0+\partial_0 \alpha(x,y,t)~~~~~~~~~~~ 
{\tilde A}_i \rightarrow {\tilde A}_i+ \partial_i \alpha(x,y,t) 
\ee 
The only term in the bulk action that it is not gauge invariant 
is the Chern-Simons term for the electromagnetic perturbations 
\be 
W_{\rm C-S}=\frac{e^2}{4\pi}\int d^3x~ \gamma(\varphi) 
\epsilon_{\mu\nu\rho} {\tilde A}_\mu \partial_\nu {\tilde A}_\rho 
\label{C-S} 
\ee 
 where $\gamma(\varphi)=\sum_{n=0}^{\infty}\Theta\left(\mu-E_n+\varphi 
\right)$ 
and $E_n=(n+\frac{1}{2})\omega_c$, where $\Theta(x)$ is the step function. 
 All the other terms in $S_{\rm eff}$ 
are gauge invariant. 
 
The variation of the Chern-Simons action Eq.\ \ref{C-S} is, 
\bea 
\delta W_{\rm C-S}=-\frac{e^2}{4\pi}\int 
d^3x&&\left\{\partial_\mu\gamma(\varphi) 
\epsilon_{\mu\nu\rho}\partial_\nu {\tilde A}_\rho-\right.\nonumber \\  
&&\left.\partial_\mu 
\left(\alpha\epsilon_{\mu\nu\rho}\partial_\nu {\tilde A}_\rho\right) 
\right\}  
\label{deltaC-S} 
\eea 
which can be split in two terms 
\be 
\delta W_{C-S} = \delta_{\cal E} W + \delta_{\cal B} W 
\ee 
 with 
\bea 
\delta_{\cal E} W&=&\frac{e^2}{2\pi}\int d^3x~ \left[ -\alpha 
\partial_i\gamma(\varphi) \epsilon_{ij}{\cal E}_j 
+ \partial_i\left\{ 
\alpha\gamma(\varphi) \epsilon_{ij}{\cal E}_j\right\} \right] 
\nonumber \\ 
&& 
\label{deltaE} \\ 
\delta_{\cal B} W&=&\frac{e^2}{2\pi}\int d^3x~  \left[-\alpha 
\partial_0\gamma(\varphi) {\cal B} 
+\partial_0\left\{ 
\alpha\gamma(\varphi){\cal B}_j\right\}\right] 
\nonumber \\ 
&& 
\label{deltaB} 
\eea 
where ${\cal E}_j=\partial_i {\tilde A}_0-\partial_0 {\tilde A}_i$  and 
${\cal B}=\epsilon_{i,j}\partial_i {\tilde A}_j$ are the electric 
and magnetic field associated to ${\tilde A}_\mu$. 
If $\gamma(\varphi)$ is a constant, then $\delta W=0$ up to boundary terms. 
However, 
\be 
\partial_\mu\gamma(\varphi)=\sum_n\delta\left(\mu-E_n+\varphi\right) 
\partial_\mu \varphi 
\ee 
Thus, in the presence of an electromagnetic field, and for 
$N$ completely filled Landau Levels, we have 
\bea 
\delta _{\cal E} W&=&-\frac{e^2}{2\pi}\int d^3x \; \alpha \; 
\delta\left(\mu-E_N+\varphi\right) \epsilon_{ij} \partial_j\varphi {\cal E}_i 
\nonumber \\ 
&+&\frac{e^2}{2\pi}\int d^3x~\partial_i\left\{ 
\alpha\gamma(\varphi) \epsilon_{ij}{\cal E}_j\right\} \label{deltaWE}\\ 
\delta_{\cal B} W&=&-\frac{e^2}{2\pi}\int d^3x \; \alpha \; 
\delta\left(\mu-E_N+\varphi\right)\partial_t\varphi {\cal B} 
\nonumber \\ 
&+&\frac{e^2}{2\pi}\int d^3x~\partial_0\left\{ 
\alpha\gamma(\varphi){\cal B}_j\right\} 
\label{deltaWB} 
\eea 
The first integral of (\ref{deltaWE}) has support on a one dimensional 
dynamical string $\partial\Omega$ defined by 
\be 
\partial\Omega:\left\{ \varphi(x,y,t)=E_N-\mu \right\} 
\label{strings} 
\ee 
The second integral is a surface term, on a 
surface that contains the string, Eq.\ \ref{strings}.  Explicitly we find 
\be 
\delta_{\cal E} W=\pm \frac{e^2}{\pi}\int_{R\times\partial\Omega} 
dsdt~ \alpha~~ \hat t\cdot \vec{\cal E}(s,t) 
\label{gaugeanomalyE} 
\ee 
where $\hat t_i=\epsilon_{ij}\partial_j\varphi/|\vec\nabla\varphi|$ is a 
unit vector tangent to the strings $\partial\Omega$. 
The $\pm$ sign is the orientation of the curve. 
 
$\delta_{\cal B}W$, of Eq.\  \ref{deltaWB}, does not vanish because 
$\varphi$ is in general a time-dependent function. 
Then, by using Eq.\ \ref{relation2} it is possible to write $\delta_{\cal 
B}W$ as a function of the  variation of the actual position of the string 
in the form 
\be 
\delta_{\cal B} W= \pm \frac{e^2}{\pi}\int_{R\times\partial\Omega} 
dsdt~ \alpha~~ 
\left(\frac{\partial \vec R}{\partial t} \cdot 
\hat n\right)~  {\cal B}(s,t) 
\label{gaugeanomalyB} 
\ee 
~From Eq.\ \ref{gaugeanomalyE} and Eq.\ \ref{gaugeanomalyB} we see that 
the divergence of the current in the bulk is 
\be 
\partial_t\rho+\partial_i J_i= \pm \frac{e^2}{\pi} \left\{\hat t\cdot \vec{\cal 
E} 
+\left(\frac{\partial \vec R}{\partial t} \cdot 
\hat n\right)~  {\cal B}\right\} 
\label{divbulk} 
\ee 
 where the ${\cal E}$ and ${\cal B}$ have support on the strings 
$\partial\Omega$ defined in Eq.\ \ref{strings}. 
 
It is not difficult to show that this divergence is canceled against the 
divergence 
of the  currents of the chiral edge states derived in Eq.\ \ref{Schiral0}. 
The induced current at the edge 
\bea 
&&J_s=\mp \frac{e^2}{\pi} \frac{\partial_s^2}{\partial_0\partial_s \mp v 
\partial_s^2} 
\left\{{\tilde A}_0(s,t) 
+\left(\frac{\partial \vec R}{\partial t} \cdot 
\hat n\right)~ {\tilde A}_n(s,t)\right\}\nonumber \\ 
&& 
\eea 
Here we have assumed the gauge ${\tilde A}_i {\hat t}_i=0$, 
where ${\tilde A}_n$ is the component of the vector potential locally normal to 
the 
strings. Evaluating the divergence of this current, we find 
\bea 
\left(\partial_0 \mp v \partial_s\right)J_s&=& 
\mp \frac{e^2}{\pi} 
\left\{\partial_s {\tilde A}_0 
+\left(\frac{\partial \vec R}{\partial t} \cdot 
\hat n\right)~ \partial_s {\tilde A}_n\right\} \nonumber \\ 
&=&\mp  \frac{e^2}{\pi} 
\left\{\hat t\cdot \vec{\cal E} 
+\left(\frac{\partial \vec R}{\partial t} \cdot 
\hat n\right)~{\cal B}(s,t)\right\} 
\nonumber \\ 
&&\label{divedge} 
\eea 
which cancels exactly Eq.\ \ref{divbulk}. In Eq.\ \ref{divedge} we 
ignored 
terms proportional to $\partial_s\left(\frac{\partial \vec R}{\partial t} \cdot 
\hat n\right)$ since they can be absorbed in a reparametrization of the curve 
$s\to f(s,t)$. 
 
%%%%%%%%%%%%%%%%%%%%%%%%%%%%%%%%%%%%%%%%% 
\section{Fluctuation propagators} 
%%%%%%%%%%%%%%%%%%%%%%%%%%%%%%%%%%%%%%%%% 
\label{sec:propagators} 
 
The propagators ${\cal F}^{\alpha\beta}$ of Eq. \ref{Dalphabeta} are the 
inverse of the fluctuation operator 
\bea 
&&\left. 
\frac{\delta^2 S_\varphi}{\delta\varphi(x')\delta\varphi(x)} 
\right|=-\left\{\left(\gamma(\varphi)+{\frac{a^2}{g}}\right) 
\nabla^2-\frac{1}{g} \right\}\delta(x-x')\nonumber \\ 
&&\label{fluctuationoperator2} 
\eea 
without the zero modes. In other words we need to evaluate the Green function 
\bea 
&& 
\left\{\left(\gamma(\varphi(x))+{\frac{a^2}{g}}\right)\nabla^2- 
\frac{1}{g}\right\} 
G(\vec x,\vec x') 
=-2\pi\delta(\vec x-\vec x') \nonumber \\ 
&&\label{GreenFunction} 
\eea 
subject to the condition that $G(\vec x,\vec x')=0$ when  
$|\vec x-\vec x'|\to\infty$. These boundary conditions automatically  
take off the zero modes since exclude any fluctuation that could 
globally translate or rotate the system.  
Notice that this propagator is {\sl static}, {\it i.\ e.\/} it is 
instantaneous. 
This feature is a consequence of the local incompressibility of the bulk 
regions. 
Eq.\ \ref{GreenFunction} is a singular partial differential 
equation due to the 
presence of the function $\gamma(\varphi(x))$. Defining the function $f(y)$ as  
\be 
f(y)=\left\{ 
\begin{array}{lcc} 
N+\frac{a^2}{g}& & n\lambda\le y < (N+\nu)\lambda \\ 
N+1+\frac{a^2}{g}& & (n+\nu)\lambda\le y < (n+1)\lambda 
\end{array} 
\right. 
\ee  
with $n$ integer and $0<\nu<1$, considering also a smooth regularization of  
$f(y)$ (i.\ e.\ finite temperature) and Fourier transforming in the coordinate  
$x$ 
\be 
G(x-x';y,y')=\int \frac{dq_x}{2\pi} e^{iq_x (x-x')} 
G(q_x;y,y') 
\ee  
we arrive to the following differential equation in $y$  
\bea 
&& 
\left[-f(y)\frac{d^2~}{dy^2}+\left(q_x^2 f(y)+\frac{1}{g}\right)\right] 
G(q_x;y,y')= \nonumber \\ 
&&(2\pi)^2\delta(y-y') 
\label{diffeq} 
\eea 
with 
\be 
G(q_x;y+\lambda,y'+\lambda)=G(q_x;y,y') 
\ee 
 
To solve Eq.\ \ref{diffeq} we adopt a recursive method. First we solve  
the equation for an arbitrary period $n$ (away form the position of the  
function $\delta$). In this case we have 
\be 
\left[-f(y)\frac{d^2~}{dy^2}+\left(q_x^2 f(y)+\frac{1}{g}\right)\right] 
F_n(y)=0 
\label{diffeqF} 
\ee 
A general solution reads,  
\be 
F^-_n(y)=A_n^- e^{K_-(y-n\lambda)}+B^-_n e^{-K_-(y-n\lambda)} 
\ee 
for $n\lambda\le y< (n+\nu)\lambda$ and  
\be 
F^+_n(y)=A_n^+ e^{K_+(y-(n+\nu)\lambda)}+B^+_n e^{-K_+(y-(n+\nu)\lambda)} 
\ee 
for $(n+\nu)\lambda\le y< (n+1)\lambda$, where we defined  
\bea 
K_-^2&=&q_x^2+\frac{1}{gN+a^2} \\ 
K_+^2&=&q_x^2+\frac{1}{g(N+1)+a^2} 
\eea 
Imposing the continuity of the function $F_n(y)$ and its derivative 
$F'_n(y)$ at the points $y=(n+\nu)\lambda$ and $y=(n+1)\lambda$ 
it is possible to find a relation between the coefficients of the solution  
in different periods. In matrix notation this relation reads 
\be 
\left( 
\begin{array}{c} 
A_{n}^- \\ 
B_{n}^- 
\end{array} 
\right) 
= M^n 
\left( 
\begin{array}{c} 
A_{0}^- \\ 
B_{0}^- 
\end{array} 
\right) 
\ee 
and  
\be 
\left( 
\begin{array}{c} 
A_{n}^+ \\ 
B_{n}^+ 
\end{array} 
\right) 
= (M')^n 
\left( 
\begin{array}{c} 
A_{0}^+ \\ 
B_{0}^+ 
\end{array} 
\right) 
\ee 
where $M$ and $M'$ are two $2\times 2$ matrices (functions of  
$K_+$ and $K_-$) with unit determinant and the subindex $0$ indicates and  
arbitrary fixed period chosen as the origin of coordinates. 
Similar equations, involving the inverse matrices $M^{-1}$ and  
$M'^{-1}$ can be found for negative values of  $n$.  
 
The propagators we are looking for are given by  
\bea 
{\cal F}^{++}_n&=&F^-_n(n\lambda)=A_n^-+B_n^-  
\label{prop++}\\ 
{\cal F}^{+-}_n&=&F^+_n((n+\nu)\lambda)=A_n^++B_n^+ 
\label{prop+-} 
\eea 
Therefore, in order to guarantee the boundary condition  
\be 
\lim_{n\to\pm\infty}{\cal F}^{\alpha\beta}_n=0 
\ee   
we choose $(A_0^-,B_0^-)$ to be an eigenvector of $M$ with eigenvalue  
$m_-<1$. For concreteness let us define the vector $(\alpha_-,\beta_-)$ 
such that 
\be 
M\left( 
\begin{array}{c} 
\alpha_- \\ 
\beta_- 
\end{array} 
\right) 
= m_- 
\left( 
\begin{array}{c} 
\alpha_- \\ 
\beta_- 
\end{array} 
\right) 
\ee 
with $m_-<1$ and $\alpha_-^2+\beta_-^2=1$. In this way we can write 
\be 
\left( 
\begin{array}{c} 
A_{n}^- \\ 
B_{n}^- 
\end{array} 
\right) 
=a\;m_-^n 
\left( 
\begin{array}{c} 
\alpha_- \\ 
\beta_-^- 
\end{array} 
\right) 
\ee 
where $a$ is an arbitrary coefficient. We can write  
a similar expression for negative $n$ 
\be 
\left( 
\begin{array}{c} 
A_{-n}^- \\ 
B_{-n}^- 
\end{array} 
\right) 
=b\;m_-^{n-1} 
\left( 
\begin{array}{c} 
\alpha_+ \\ 
\beta_+ 
\end{array} 
\right) 
\ee 
where $(\alpha_+,\beta_+)$ is a unit eigenvector of the matrix $M^{-1}$. 
(the eigenvalue is the same due to ${\rm det}M=1$). 
 
Finally, let us suppose that the $\delta$-function has support in $y'=0$. 
We can determine the two unknown coefficients $a$ and $b$ by  
asking continuity of the function  and discontinuity of the derivative 
at the origin. 
\bea 
F_0^-(0)-F_{-1}^+(0)&=&0 \\ 
\left.\frac{d F_0^-}{dy}\right|_{y=0}- 
\left.\frac{d F_{-1}^+}{dy}\right|_{y=0}&=&-\frac{(2\pi)^2}{f(0)} 
\eea 
where we choose the regularization of $f(y)$ such that  
$f(0)=N+\frac{a^2}{g}+\frac{1}{2}$. 
 
Following this tedious by direct algebra it is possible to  
exactly determine the propagators of Eq.\ \ref{prop++} and Eq.\   
\ref{prop+-}. Although the result is a complicated expression, 
it can be cast in a simpler form considering the limit $Ng+a^2>>g$ and  
$\nu=1/2$. In this limit, the difference $\delta=K_+-K_-$ is an infinitesimal 
quantity and the propagators to leading order in $\delta$ read 
\be 
{\cal F}^{++}_n(q_x)=2\pi^2\frac{e^{-|n| K_-\lambda}} 
{K_-(N+\frac{a^2}{g})} 
\label{Delta++n} 
\ee 
\bea 
&& 
{\cal F}^{+-}_n(q_x)=2\pi^2\frac{e^{- K_-\lambda/2}} 
{K_-(N+\frac{a^2}{g})} 
\left\{ 
\begin{array}{lcl} 
e^{-n K_-\lambda} & & n\ge 0 \\ 
 & & \\ 
e^{(n+1) K_-\lambda} & & n< 0  
\end{array} 
\right.  \nonumber \\ 
&&\label{Delta+-n} 
\eea  
To obtain the propagators in momentum $q_y$ space 
we make the following Fourier transformation 
\be 
{\cal F}^{\alpha\beta}(q_x,q_y)=\sum_{n=-\infty}^{\infty}  
{\cal F}_n^{\alpha\beta}(q_x) e^{i q_y n} 
\ee 
It is straightforward to obtain  
\bea 
{\cal F}^{++}&=& 
16\pi^2\; v\; 
\frac{\sinh(K_-\lambda)}{K_-\lambda} 
\frac{1}  
{\cosh(K_-\lambda)-\cos q_y} 
\nonumber \\ 
&& 
\label{ADelta++qxqy} 
\eea 
\bea 
{\cal F}^{+-}&=& 
16\pi^2\; v\;   
\frac{\sinh(K_-\lambda/2)}{K_-} 
\frac{(1+e^{-iq_y})} 
{\cosh(K_-\lambda)-\cos q_y} 
\nonumber \\ 
&& 
\label{ADelta+-qxqy} 
\eea 
where in this approximation  
\be 
v=\frac{1}{2}\frac{g^{3/2}}{\sqrt{gN+a^2}} 
\label{v} 
\ee 
In the long wavelength limit, $q_x \to 0$ and $q_y \to 0$, the kernels ${\cal 
F}_{\alpha \beta}(q)$ take the finite limiting values 
\bea 
{\cal F}^{+ +}(0)&=&\frac{16 \pi^2 {v}}{\lambda \xi_N} \coth\left(\lambda 
\xi_N/2\right) \\ 
{\cal F}^{+ -}(0)&=&\frac{16 \pi^2 {v}}{\lambda \xi_N}  
\frac{1}{\sinh\left(\lambda \xi_N/2\right)} \\ 
\label{eq:F0} 
\eea 
which are simple smooth functions of the coupling constant. 
 
%%%%%%%%%%%%%%%%%%%%%%%%%%%%%%%%%%%%%%%%%%%%%%%%%%%%%%%%%%%%% 
\section{Elastic constants} 
\label{sec:constants} 
%%%%%%%%%%%%%%%%%%%%%%%%%%%%%%%%%%%%%%%%%%%%%%%%%%%%%%%%%%%%% 
 
In this appendix we show some details of the calculation of the  
constants $Q$, $\kappa_\perp$ and $\kappa_\parallel$ that enter the action of  
Eq.\ \ref{Susmectic}. 
 
The elastic constants $Q$ and $\kappa_\perp$ are obtained by replacing in  
$S_{\varphi}$ (Eq.\ \ref{Svarphi}) the deformed saddle-point solution  
$\varphi(u)$ (Eq.\ \ref{deformedsp}). The main contribution to $Q$ comes from  
the  
first term of Eq.\  \ref{Svarphi},  
 
\be 
\int \frac{d^3x}{4\pi}\;\frac{3 \gamma^2}{8}(\partial^2_x\varphi(u))^2   
=\int \frac{d^3x}{4\pi}\;\frac{3 \gamma^2 (\varphi')^2}{8} 
(\partial^2_x u)^2  
\label{Qint} 
\ee 
where $\varphi'$ is the derivative of $\phi$.  
While the term $\partial^2_x u$ in Eq. \ref{Qint} is a slowly varying  
function of $y$ in a scale long with respect to the stripe period $\lambda$,  
the expression $\gamma^2 (\varphi')^2$ is a rapidly varying function within  
a period $\lambda$.  
Therefore, at long distances,  we can safely take the mean value of the last  
expression over one period, obtaining 
\be 
\int \frac{d^3x}{4\pi}\;\frac{3 \gamma^2}{8}(\partial^2_x\varphi(u))^2   
=\int d^3x\; Q\;\; 
(\partial^2_x u)^2  
\label{Qint2} 
\ee  
where we have defined  
 
\be 
Q=\frac{3}{32 \pi \lambda}\int_0^\lambda dy~\left( 
\gamma(\varphi)\frac{\partial\varphi(y)}{\partial y} \right)^2 
\label{K1} 
\ee 
In momentum space, this corse-graining procedure is equivalent to take the zero  
momentum limit of the Fourier transform of $\gamma(\varphi)\phi'$. We simply  
obtain  
\be 
Q=\frac{3}{32\pi} N^2 v^2 
\ee  
In the limit $Ng+a^2>>g$ we can use Eq.\ref{v} for the velocity obtaining  
\be 
Q=\frac{3}{128\pi}\;\frac{N^2g^2}{gN+a^2} 
\ee 
In Eq.\ \ref{eq:constants} we show the value of $Q$ for a very short ranged  
potential $a\to 0$.  
 
The compressibility $\kappa_\perp$ is nothing but the  
energy density per period of the saddle-point configuration  
calculated in Appendix \ref{sec:mean-field} (see Eq.\  \ref{totalenergyN}). 
To formally obtain this expression we proceed as follows: first we  
substitute $\varphi$ by the deformed saddle-point $\varphi(u)$ in  
Eq.\ \ref{Svarphi}. Then we performed the change of variables $y'=y\alpha+u$ 
obtaining to leading order in the derivatives of $u$ 
\bea 
\lefteqn{ 
\int d^3x {\cal L}(\varphi(u(x,y)))\approx 
\int d^3x {\cal L}(\varphi(y))}\nonumber \\ 
 &+&\frac{1}{2}\int d^3x {\cal L}(\varphi(y))  
\left(\frac{\partial u}{\partial y} \right)^2+\ldots 
\nonumber \\ 
\label{jac} 
\eea 
Again, in the last integral, the factor ${\cal L}(\varphi)$ is a periodic  
function  
of  
$y$ with period $\lambda$, while $\partial_y u$ is a slowly varying function of  
$y$. Therefore we can take the mean value of ${\cal L}(\varphi)$ on a period  
defining in this way 
 
\bea 
\kappa_\perp&=&\frac{1}{2\lambda}\int_0^\lambda dy~\left\{ 
\left[\gamma+\frac{a^2}{2g}\right]\left(\frac{\partial\varphi(y)}{\partial y}  
\right)^2 
+\gamma \varphi \right.\nonumber \\ 
&+&\left.\left( 
\frac{1}{2g}+\frac{v+{\cal F}^{++}-{\cal R}e{\cal F}^{+-}} 
{(v+{\cal F}^{++})^2-|{\cal F}^{+-}|^2}\right)\varphi^2 \right\}  
\label{B} 
\eea 
where $\varphi$ is the undeformed saddle-point solution. 
The kernels ${\cal F}^{++}$ and ${\cal F}^{+-}$ come from the contribution of  
the chiral modes to the action and are 
given in Appendix \ref{fluctuations}. Notice that only the ${\vec q} \to 0$ 
limit of this kernels is important here.  In the limit $gN+a^2>>g$ and $g<<1$  
this expression is given by,  
\be 
\kappa_\perp=\frac{5}{4} N^{2} g^{1/2}\sqrt{gN+a^2} 
\ee 
The dynamical term in Eq. \ref{Susmectic} comes form the Gaussian integral of  
the chiral modes and is given by Eq.\ \ref{integration}. We can rewrite it  
more  explicitly as  
\bea 
-\frac{1}{2}\sum_{nm}\int d^2xd^2x'\; 
\varphi'_n \varphi'_m \; \Pi_{mn} \; 
\partial_x u_m \partial_x u_n   
\eea  
where 
\be 
\Pi_{mn}=\left\{\pi_{++}^{-1}+\pi_{--}^{-1}-\pi_{+-}^{-1}-\pi_{-+}^{-1} 
\right\}_{mn} 
\ee 
Calculating the inverse matrix $\pi^{-1}$, subtracting the static part, and  
taking  the continuum limit as in the preceding cases we finally find the  
dynamical contribution to the  
Lagrangian (in momentum space),  
 
\be 
{\cal L}_{\rm dyn}=-\frac{1}{2} K \frac{\omega^2}{\omega^2-\bar v^2 q_x^2} 
\ee 
where the renormalized velocity is given by  
\be 
\bar v^2= (v+{\cal F}^{++})^2-|{\cal F}^{+-}|^2 
\ee 
and the constant $K$ can be calculated from  
\be 
K=\frac{1}{\lambda}\int_0^\lambda dy   
\frac{v+{\cal F}^{++}-{\cal R}e {\cal F}^{+-}}{((v+{\cal F^{++}})^2- 
|{\cal F}^{+-}|^2)} \left(\frac{\partial \varphi}{\partial y}\right)^2 
\label{K2} 
\ee 
~From Eq.\ \ref{integration} we have that  
\be 
\kappa_\parallel=\frac{\bar v^2}{K \lambda} 
\ee 
Using the expression for ${\cal F}^{\alpha\beta}$ given in Appendix  
\ref{sec:propagators} we find in the limit $Ng+a^2>>1$ and $g<<1$  
\be 
\kappa_\parallel=\frac{\pi^2}{16}\; g^{-3} N^{-3/2} 
\ee

%%%%%%%%%%%%%%%%%%%%%%%%%%%%%%%%%%%%%%%%%%%%%%%%%%%%%%%%%%%% 

\end{document}